\documentclass[aps,prd,nofootinbib,amsmath,amssymb,superscriptaddress,twocolumn,10pt]{revtex4}

\pdfoutput=1

\usepackage{graphicx}
\usepackage{dcolumn}
\usepackage{bm}
\usepackage{amssymb}
\usepackage{latexsym}
\usepackage{booktabs}
\usepackage{amsmath}
\usepackage{multirow}
\usepackage[colorlinks=true, linkcolor=red, citecolor=blue]{hyperref}

\newcommand{\be}{\begin{equation}}
\newcommand{\ee}{\end{equation}}
\newcommand{\bq}{\begin{eqnarray}}
\newcommand{\eq}{\end{eqnarray}}

\begin{document}

\title{Comparison of dark energy models after Planck 2015}

\author{Yue-Yao Xu}
\affiliation{Department of Physics, College of Sciences, Northeastern University, Shenyang
110004, China}
\author{Xin Zhang\footnote{Corresponding author}}
\email{zhangxin@mail.neu.edu.cn} \affiliation{Department of Physics, College of Sciences,
Northeastern University, Shenyang 110004, China}
\affiliation{Center for High Energy Physics, Peking University, Beijing 100080, China}

\begin{abstract}
We make a comparison for ten typical, popular dark energy models according to their capabilities of fitting the current observational data. The observational data we use in this work include the JLA sample of type Ia supernovae observation, the Planck 2015 distance priors of cosmic microwave background observation, the baryon acoustic oscillations measurements, and the direct measurement of the Hubble constant. Since the models have different numbers of parameters, in order to make a fair comparison, we employ the Akaike and Bayesian information criteria to assess the worth of the models. The analysis results show that, according to the capability of explaining observations, the cosmological constant model is still the best one among all the dark energy models. The generalized Chaplygin gas model, the constant $w$ model, and the $\alpha$ dark energy model are worse than the cosmological constant model, but still are good models compared to others. The holographic dark energy model, the new generalized Chaplygin gas model, and the Chevalliear-Polarski-Linder model can still fit the current observations well, but from an economically feasible perspective, they are not so good. The new agegraphic dark energy model, the Dvali-Gabadadze-Porrati model, and the Ricci dark energy model are excluded by the current observations.
\end{abstract}
\pacs{95.36.+x, 98.80.Es, 98.80.-k} \maketitle

\section{Introduction}
\label{sec:intro}

The current astronomical observations have indicated that the universe is undergoing an accelerated expansion~\cite{Riess:1998cb,Perlmutter:1998np,Tegmark:2003ud,Eisenstein:2005su,Spergel:2003cb}, for which a natural explanation is that the universe is currently dominated by dark energy (DE) that has negative pressure. The study of the nature of dark energy has become one of the most important issues in the field of fundamental physics~\cite{Sahni:1999gb,Padmanabhan:2002ji,Bean:2005ru,Copeland:2006wr,Sahni:2006rde,Kamionkowski:2007wv,Frieman:2008sn,Li:2011sd,Bamba:2012cp}. But, hitherto, we still know little about the physical nature of dark energy. The simplest candidate for dark energy is the Einstein's cosmological constant, $\Lambda$, which is physically equivalent to the quantum vacuum energy. For $\Lambda$, one has the equation of state $p_\Lambda=-\rho_\Lambda$. The cosmological model with $\Lambda$ and cold dark matter (CDM) is usually called the $\Lambda$CDM model, which can explain the current various astronomical observations quite well. But the cosmological constant has always been facing the severe theoretical challenges, such as the fine-tuning and coincidence problems.

There also exist many other possible theoretical candidates for dark energy. For example, a spatially homogeneous, slowly rolling scalar field can also provide a negative pressure, driving the cosmic acceleration. Such a light scalar field is usually called ``quintessence'' \cite{Zlatev:1998tr,Steinhardt:1999nw,Zhang:2005rg,Zhang:2005rj}, which provides a possible mechanism for dynamical dark energy. More generally, one can phenomenologically characterize the property of dynamical dark energy through parametrizing $w$ of its equation of state (EoS) $p_{\rm de}=w\rho_{\rm de}$, where $w$ is usually called the EoS parameter of dark energy. For example, the simplest parametrization model corresponds to the case of $w={\rm constant}$, and this cosmological model is sometimes called the $w$CDM model. A more physical and realistic situation is that $w$ is time variable, which is often probed by the so-called Chevalliear-Polarski-Linder (CPL) parametrization \cite{MD:2001,EV:2003}, $w(a)=w_0+w_a(1-a)$. For other popular parametrizations, see, e.g., \cite{Huterer:2000mj,Wetterich:2004pv,Jassal:2004ej,Upadhye:2004hh,Xia:2004rw,Linder:2006sv,Lazkoz:2010gz,Ma:2011nc,Li:2011dr,Li:2012vn,Li:2012via}.

Some dynamical dark energy models are built based on deep theoretical considerations. For example, the holographic dark energy (HDE) model has a quantum gravity origin, which is constructed by considering the holographic principle of quantum gravity theory in a quantum effective field theory~\cite{A.G:1999wv,M.Li:2004wv}. The HDE model can naturally explain the fine-tuning and coincidence problems \cite{M.Li:2004wv} and can also fit the observational data well \cite{hde1,hde2,hde3,hde4,hde5,hde6,hde7,Li:2009zs,Li:2009bn,Wang:2012uf,Wang:2013zca,Zhang:2015rha,Feng:2016djj,He:2016rvp}. Its theoretical variants, the new agegraphic dark energy (NADE) model \cite{HWRG:2008} and the Ricci dark energy (RDE) model \cite{CGFQ:2009}, have also attracted lots of attention. In addition, the Chaplygin gas model \cite{Kamenshchik:2001cp} is motivated by braneworld scenario, which is claimed to be a scheme for unifying dark energy and dark matter. To fit the observational data in a better way, its theoretical variants, the generalized Chaplygin gas (GCG) model \cite{MCB:2002} and the new generalized Chaplygin gas (NGCG) model \cite{Zhang:2004gc}, have also been put forward. Moreover, actually, the cosmic acceleration can also be explained by the modified gravity (MG) theory, i.e., the theory in which the gravity rule deviates from the Einstein general relativity (GR) on the cosmological scales. The MG theory can yield ``effective dark energy'' models mimicking the real dark energy at the background cosmology level.\footnote{Usually, the growth of linear matter perturbations in the MG models is distinctly different from that in the DE models within GR.} Thus, if we omit the issue of growth of structure, we may also consider such effective dark energy models. A typical example of this type is the Dvali-Gabadadze-Porrati (DGP) model \cite{a:2000}, which arises from a class of braneworld theories in which the gravity leaks out into the bulk at large distances, leading to the accelerated expansion of the universe. Also, its theoretical variant, the $\alpha$DE model \cite{c:2003}, can fit the observational data much better.

Facing so many competing dark energy models, the most important mission is to find which one on earth is the right dark energy model. But this is too difficult. A more realistic mission is to select which ones are better than others in explaining the various observational data. Undoubtedly, the right dark energy model can certainly fit all the astronomical observations well. The Planck satellite mission has released the most accurate data of cosmic microwave background (CMB) anisotropies, which, combining with other astrophysical observations, favor the base $\Lambda$CDM model \cite{Ade:2013zuv,Ade:2015xua}. But it is still necessary to make a comparison for the various typical dark energy models by using the Planck 2015 data and other astronomical data to select which ones are good models in fitting the current data. Such a comparison can also help us to discriminate which models are actually excluded by the current observations.

We use the  $\chi^2$ statistic to do the cosmological fits, but we cannot fairly compare different models by comparing their $\chi_{\rm min}^2$ values because they have different numbers of parameters. It is obvious that a model with more free parameters would tend to have a lower $\chi_{\rm min}^2$. Therefore, in this paper, we use the information criteria (IC) including the Akaike information criterion (AIC)~\cite{H.Akaike:1974} and the Bayesian information criterion (BIC)~\cite{G.schwarz:1978} to make a comparison for different dark energy models. The IC method has sufficiently taken the factor of number of parameters into account. Of course, we will use the uniform data combination of various astronomical observations in the model comparison. In this work, we choose ten typical, popular dark energy models to make a uniform, fair comparison. We will find that, compared to the early study~\cite{Zhang.xin:2010}, in the post-Planck era we are now truly capable of discriminating different dark energy models.

The paper is organized as follows. In Sect. \ref{sec.2} we introduce the method of information criteria and how it works in comparing competing models. In Sect. \ref{sec.3} we describe the current observational data used in this paper. In Sect. \ref{sec.4} we describe the ten typical, popular dark energy models chosen in this work and give their fitting results. We discuss the results of model comparison and give the conclusion in Sect. \ref{sec.5}.

\section{Methodology}\label{sec.2}

We use the $\chi^2$ statistic to fit the cosmological models to observational data. The $\chi^2$ function is given by
\begin{equation}
\chi^2_{\xi}=\frac{(\xi_{\rm {th}}-\xi_{\rm {obs}})^{2}}{\sigma^{2}_{\xi}},
\end{equation}
where $\xi_{\rm{obs}}$ is the experimentally measured value, $\xi_{\rm{th}}$ is the theoretically predicted value, and $\sigma_{\xi}$ is the standard deviation.
The total $\chi^2$ is the sum of all $\chi^2_{\xi}$,
\begin{equation}
\chi^2=\sum\limits_{\xi}{\chi^{2}_{\xi}}.
\end{equation}

In this paper, we use the observational data including the type Ia supernova (SN) data from the ``joint light-curve analysis" (JLA) compilation, the CMB data from the Planck 2015 mission, the baryon acoustic oscillation (BAO) data from the 6dFGS, SDSS-DR7, and BOSS-DR11 surveys, and the direct measurement of the Hubble constant $H_{0}$ from the Hubble Space Telescope (HST). So the total $\chi^{2}$ is written as
\begin{equation}
\chi^2=\chi^{2}_{\rm{SN}}+\chi^{2}_{\rm{CMB}}+\chi^{2}_{\rm{BAO}}+\chi^{2}_{H_{0}}.
\end{equation}

We cannot make a fair comparison for different dark energy models by directly comparing their values of $\chi^2$, because they have different numbers of parameters. Obviously, a model with more parameters is more prone to have a lower value of $\chi^2$. Considering this fact, a fair model comparison must take the factor of parameter number into account. In this work, we apply the IC method to do the analysis. We employ the AIC~\cite{H.Akaike:1974} and BIC~\cite{G.schwarz:1978} to do the model comparison, which are rather popular among the information criteria.

The AIC~\cite{H.Akaike:1974} is defined as
\begin{equation}
{\rm AIC}=-2\ln{\mathcal{L}_{\rm{max}}}+2k,
\end{equation}
where $\mathcal{L}_{\rm max}$ is the maximum likelihood and $k$ is the number of parameters. It should be noted that, for Gaussian errors, $\chi^{2}_{\rm{min}}=-2\ln{\mathcal{L}_{\rm{max}}}$. In practice, we do not care about the absolute value of the criterion, and we actually pay more attention to the relative values between different models, i.e., $\Delta {\rm AIC}=\Delta\chi^{2}_{\rm{min}}+2\Delta k$. A model with a lower AIC value is more favored by data. Among many models, one can choose the model with minimal value of AIC as a reference model. Roughly speaking, the models with $0<\Delta {\rm AIC}<2$ have substantial support, the models with $4<\Delta {\rm AIC}<7$ have considerably less support, and the models with $\Delta {\rm AIC}>10$ have essentially no support, with respect to the reference model.

The BIC~\cite{G.schwarz:1978}, also known as the Schwarz information criterion, is given by
\begin{equation}
{\rm BIC}=-2\ln{\mathcal{L}_{\rm{max}}}+k\ln{N},
\end{equation}
where $N $ is the number of data points used in the fit. The same as AIC, the relative value between different models can be written as $\Delta{\rm BIC}=\Delta\chi^{2}_{\rm{min}}+\Delta k\ln{N}$. A difference in $\Delta{\rm BIC}$ of 2 is considerable positive evidence against the model with higher BIC, while a $\Delta{\rm BIC}$ of 6 is considered to be strong evidence. The model comparison needs to choose a well justified single model, so in our work, the same as Refs. \cite{Zhang.xin:2010,M. Szydlowski:2006,b:777}, we use the $\Lambda$CDM model to play this role. Thus, the values of $\Delta{\rm AIC}$ and $\Delta{\rm BIC}$ are measured with respect to the $\Lambda$CDM model.

The AIC only considers the factor of parameter number but does not consider the factor of data point number. Thus, once the data point number is large, the result would be in favor of the model with more parameters. In order to further penalize models with more parameters, the BIC also takes the number of data points into account. Considering both AIC and BIC could provide us with more reasonable perspective to the model comparison.

\section {The observational data}\label{sec.3}

We use the combination of current various observational data to constrain the dark energy models chosen in this paper. Using the fitting results, we make a comparison for these dark energy models and select the good ones among the models. In this section, we describe the cosmological observations used in this paper. Since the smooth dark energy affects the growth of structure only through the expansion history of the universe, different smooth dark energy models yield almost the same growth history of structure. Thus, in this paper, we only consider the observational data of expansion history, i.e., those describing the distance-redshift relations. Specifically, we use the JLA SN data, the Planck CMB distance prior data, the BAO data, and the $H_0$ measurement.

\subsection{The SN data}

We use the JLA compilation of type Ia supernovae~\cite{M.Betoule:2014}. The JLA compilation is from a joint analysis of type Ia supernova observations in the redshift range of $z\in [0.01,1.30]$. It consists of 740 Ia supernovae, which collects several low-redshift samples, obtained from three seasons from SDSS-II, three years from SNLS, and a few high-redshift samples from the HST. According to the observational point of view, we can get the distance modulus of a SN Ia from its light curve through the empirical linear relation~\cite{M.Betoule:2014},
\begin{equation}
\hat{\mu}=m^{\ast}_{\rm{B}}-(M_{\rm{B}}-\alpha \times X_{1}+\beta \times C),
\end{equation}
where $m^{\ast}_{\rm{B}}$ is the observed peak magnitude in the rest frame B band, $M_{\rm{B}}$ is the absolute magnitude which depends on the host galaxy properties complexly, $X_{1}$ is the time stretching of the light curve, and $C$ is the supernova color at maximum brightness. For the JLA sample, the luminosity distance $d_{\rm{L}}$ of a supernova can be given by
\begin{equation}
d_{{\rm L}}(z_{\rm{hel}},z_{\rm{cmb}})=\frac{1+z_{\rm{hel}}}{H_{0}} \int_{0}^{z_{\rm{cmb}}} \frac{dz'}{E(z')},
\end{equation}
where $z_{\rm{cmb}}$ and $z_{\rm{hel}}$ denote the CMB frame and heliocentric redshifts, respectively, $H_{0}=100h$ km s$^{-1}$ Mpc$^{-1}$ is the Hubble constant, $E(z)=H(z)/H_0$ is given by a specific cosmological model. The $\chi^{2}$ function for JLA SN observation is written as
\begin{equation}
\chi^{2}_{\rm{SN}}=(\hat{\mu}-\mu_{\rm{th}})^{\dagger}C_{\rm SN}^{-1}(\hat{\mu}-\mu_{\rm{th}}),
\end{equation}
where $C_{\rm SN}$ is the covariance matrix of the JLA SN observation and $\mu_{\rm{th}}$ denotes the theoretical distance modulus,                                                                                                 \begin{equation}
\mu_{\rm{th}}=5\log_{10}\frac{d_{\rm{L}}}{10\rm{pc}}.
\end{equation}

\subsection{The CMB data}

The CMB data alone cannot constrain dark energy well, because the main effects constraining dark energy in the CMB anisotropy spectrum come from a angular diameter distance to the decoupling epoch $z\simeq 1100$ and the late integrated Sachs-Wolfe (ISW) effect. The late ISW effect cannot be accurately measured currently, and so the only important information for constraining dark energy in the CMB data actually comes from the angular diameter distance to the last scattering surface, which is important because it provides a unique high-redshift ($z\simeq 1100$) measurement in the multiple-redshift joint constraint. In this work, we focus on the smooth dark energy models, in which dark energy mainly affects the expansion history of the universe. Thus, for an economical reason, we do not use the full data of the CMB anisotropies, but decide to use the compressed information of CMB, i.e., the CMB distance priors.

We use the ``Planck distance priors" from the Planck 2015 data~\cite{Planck Collaboration:2015}. The distance priors contain the shift parameter $R$, the ``acoustic scale" $\ell_{\rm{A}}$, and the baryon density $\omega_{\rm{b}}\equiv\Omega_{\rm{b}}h^{2}$,
\begin{equation}
R\equiv\sqrt{\Omega_{\rm{m}}H^{2}_{0}}(1+z_{\ast})D_{\rm{A}}(z_{\ast}),
\end{equation}
and
\begin{equation}
\ell_{\rm{A}}\equiv(1+z_{\ast})\frac{\pi D_{\rm{A}}(z_{\ast})}{r_{\rm{s}}(z_{\ast})},\label{la}
\end{equation}
where $\Omega_{\rm{m}}$ is the present-day fractional energy density of matter, $D_{\rm{A}}(z_{\ast})$ is the proper angular diameter distance at the redshift of the decoupling epoch of photons $z_{\ast}$.
Because we consider a flat universe, $D_{\rm{A}}$ can be expressed as
\begin{equation}
D_{\rm{A}}(z)=\frac{1}{H_{0}(1+z)}\int_{0}^{z}\frac{dz'}{E(z')}.\label{DA}
\end{equation}
In Eq.~(\ref{la}), $r_{\rm{s}}(z)$ is the comoving sound horizon at $z$,
\begin{equation}
r_{\rm{s}}(z)=H_{0}^{-1}\int_{0}^{a}\frac{da'}{a'E(a')\sqrt{3(1+\overline{R_{\rm{b}}}a')}},\label{rs}
\end{equation}
where $\overline{R_{\rm{b}}}a=3\rho_{\rm{b}}/(4\rho_{\gamma})$. It should be noted that $\rho_{\rm{b}}$ is the baryon energy density, $\rho_{\gamma}$ is the photon energy density, and both of them are the present-day energy densities. Thus we have $\overline{R_{\rm{b}}}=31500\Omega_{\rm{b}}h^{2}(T_{\rm{cmb}}/2.7{\rm K})^{-4}$. We take $T_{\rm{cmb}}=2.7255$ K. $z_{\ast}$ is given by the fitting formula~\cite{Hu:1995en},
\begin{equation}
z_{\ast}=1048[1+0.00124(\Omega_{{\rm b}}h^{2})^{-0.738}][1+g_{1}(\Omega_{{\rm m}}h^{2})^{g_{2}}],
\end{equation}
where
\begin{equation}
g_{1}=\frac{0.0783(\Omega_{\rm{b}}h^{2})^{-0.238}}{1+39.5(\Omega_{\rm{b}}h^{2})^{-0.76}}, \;  g_{2}=\frac{0.560}{1+21.1(\Omega_{\rm{b}}h^{2})^{1.81}}.
\end{equation}
Using the Planck TT+LowP data, the three quantities are obtained: $R=1.7488\pm0.0074$, $\ell_{\rm{A}}=301.76\pm0.14$, and $\Omega_{\rm{b}}h^{2}=0.02228\pm0.00023$. The inverse covariance matrix for them, ${\rm Cov}^{-1}_{\rm CMB}$, can be found in Ref.~\cite{Planck Collaboration:2015}. The $\chi^{2}$ function for CMB is
\begin{equation}
\chi^{2}_{\rm{CMB}}=\Delta p_{i}[{\rm Cov}^{-1}_{\rm{CMB}}(p_{i},p_{j})]\Delta p_{j}, \quad \Delta p_{i}=p_{i}^{\rm{th}}-p_{i}^{\rm{obs}},
\end{equation}
where $p_{1}=\ell_{\rm{A}}$, $p_{2}=R$, and $p_{3}=\omega_{\rm{b}}$.

\subsection{The BAO data}

The BAO signals can be used to measure not only the angular diameter distance $D_{\rm{A}}(z)$
through the clustering perpendicular to the line of sight, but also the expansion rate of the universe $H(z)$ by the clustering along the line of sight. We can use the BAO measurements to get the ratio of the effective distance measure $D_{{\rm V}}(z)$ and the comoving sound horizon size $r_{\rm{s}}(z_{\rm{d}})$. The spherical average gives us the expression of $D_{\rm{V}}(z)$,
\begin{equation}
D_{\rm{V}}(z)\equiv\left[(1+z)^{2}D^{2}_{\rm{A}}(z)\frac{z}{H(z)}\right]^{1/3}.
\end{equation}
The comoving sound horizon size $r_{\rm{s}}(z_{\rm{d}})$ is given by Eq.~(\ref{la}), where $z_{\rm{d}}$ is the redshift of the drag epoch, and its fitting formula is given by~\cite{Eisenstein:1997ik}
\begin{equation}
z_{\rm{d}}=\frac{1291(\Omega_{\rm{m}}h^2)^{0.251}}{1+0.659(\Omega_{\rm{m}}h^2)^{0.828}}[1+b_1(\Omega_{\rm{b}}h^2)^{b_2}],
\end{equation}
where
\begin{equation}
\begin{gathered}
b_1=0.313(\Omega_{\rm{m}}h^2)^{-0.419}[1+0.607(\Omega_{\rm{m}}h^2)^{0.674}],\\
b_2=0.238(\Omega_{\rm{m}}h^2)^{0.223}.
\end{gathered}
\end{equation}
We use four BAO data points: $r_{\rm{s}}(z_{\rm{d}})/D_{\rm{V}}(0.106)=0.336\pm0.015$ from the 6dF Galaxy Survey~\cite{Beutler.F:2011}, $D_{\rm{V}}(0.15)=(664\pm25{\rm Mpc})(r_{\rm{d}}/r_{\rm{d, fid}})$ from the SDSS-DR7~\cite{Ross.A.J:2014}, $D_{\rm{V}}(0.32)=(1264\pm25{\rm Mpc})(r_{\rm{d}}/r_{\rm{d, fid}})$ and $D_{{\rm V}}(0.57)=(2056\pm20{\rm Mpc})(r_{\rm{d}}/r_{\rm{d, fid}})$ from the BOSS-DR11~\cite{Anderson.L:2014}. Note that in this paper we do not use the WiggleZ data because the WiggleZ volume partially overlaps with the BOSS-CMASS sample, and the WiggleZ data are correlated with each other but we could not quantify this correlation~\cite{Kazin:2014qga}. The $\chi^{2}$ function for BAO is
\begin{equation}
\chi^{2}_{\rm{BAO}}=\Delta p_{i}[{\rm Cov}^{-1}_{\rm{BAO}}(p_{i},p_{j})]\Delta p_{j}, \; \Delta p_{i}=p_{i}^{\rm{th}}-p_{i}^{\rm{obs}}.
\end{equation}
Since we do not include the WiggleZ data in the analysis, the inverse covariant matrix ${\rm Cov}^{-1}_{\rm{CMB}}$ is a unit matrix in this case.

{\subsection{The $H_{0}$ measurement}

We use the result of direct measurement of the Hubble constant, given by Efstathiou~\cite{G.Efstathiou}, $H_{0}=70.6\pm3.3$  km s$^{-1}$ Mpc$^{-1}$, which is derived from a re-analysis of Cepheid data of Riess et al.~\cite{A. G. Riess:2011} by using the revised geometric maser distance to NGC 4258. The $\chi^{2}$ function for the $H_{0}$ measurement is
\begin{equation}
\chi^{2}_{H_{0}}=\left(\frac{h-0.706}{0.033}\right)^{2}.
\end{equation}

Note that the various observations used in this paper are consistent with each other. More recently, Riess et al. \cite{Riess:2016jrr} obtained a very accurate measurement of the Hubble constant (a 2.4\% determination), $H_0=73.00\pm 1.75$ km s$^{-1}$ Mpc$^{-1}$. But this measurement is in tension with the Planck data. To relieve the tension, one might need to consider the extra relativistic degrees of freedom, i.e., the additional parameter $N_{\rm eff}$. In addition, the measurements from the growth of structure, such as the weak lensing, the galaxy cluster counts, and the redshift space distortions, also seem to be in tension with the Planck data \cite{Ade:2013zuv}. Considering massive neutrinos as a hot dark matter component might help to relieve this type of tension. Synthetically, the consideration of light sterile neutrinos is likely to be a key to a new concordance model of cosmology \cite{Zhang:2014dxk,Dvorkin:2014lea}. But this is not the issue of this paper. In this work, we mainly consider the smooth dark energy models, and thus the combination of the SN, CMB, BAO, and $H_0$ data is sufficient for our mission. The various observations described in this paper are consistent.

\section {Dark energy models}\label{sec.4}

In this section, we briefly describe the dark energy models that we choose to analyze in this paper and discuss the basic characteristics of these models. At the same time, we give the fitting results of these models by using the observational data given in the above section.

In a spatially flat FRW universe ($\Omega_{\rm{k}}=0$), the Friedmann equation can be written as
\begin{equation}
3M_{\rm{pl}}^{2}H^{2}=\rho_{\rm{m}}(1+z)^{3}+\rho_{\rm{r}}(1+z)^{4}+\rho_{\rm{de}}(0)f(z),\label{DEM}
\end{equation}
where $M_{\rm{pl}}\equiv\frac{1}{\sqrt{8 \pi G}}$ is the reduced Planck mass, $\rho_{\rm{m}}$, $\rho_{\rm{r}}$, and $\rho_{\rm{de}}(0)$ are the present-day densities of dust matter, radiation, and dark energy, respectively. It should be noted that $f(z)\equiv\frac{\rho_{\rm{de}}(z)}{\rho_{\rm{de}}(0)}$, which is given by the specific dark energy models. From Eq.~(\ref{DEM}), we have
\begin{equation}
\begin{aligned}
E(z)^{2}\equiv\ \left(\frac{H(z)}{H_0}\right)^{2}&=\Omega_{\rm{m}}(1+z)^{3}+\Omega_{\rm{r}}(1+z)^{4}\\
&+(1-\Omega_{\rm{m}}-\Omega_{\rm{r}})f(z).
\end{aligned}
\end{equation}
Here in our work the radiation density parameter $\Omega_{\rm{r}}$ is given by 
\begin{equation}
\Omega_{{\rm r}}=\Omega_{{\rm m}} / (1+z_{\rm eq}),
\end{equation}
where $z_{\rm eq}=2.5\times 10^4 \Omega_{{\rm m}} h^2 (T_{\rm cmb}/2.7\,{\rm K})^{-4}$. 

In this paper, we choose ten typical, popular dark energy models to analyze. We constrain these models with the same observational data, and then we make a comparison for them. From the analysis, we will know which model is the best one in fitting the current data and which models are excluded by the current data. We divide these models into five classes:  \\
   (a) Cosmological constant model. \\
   (b) Dark energy models with equation of state parameterized. \\
   (c) Chaplygin gas models.  \\
   (d) Holographic dark energy models. \\
   (e) Dvali-Gabadadze-Porrati (DGP) braneworld and related models.

Here we ignore the exiguous difference between DE and MG models because we only consider the aspect of acceleration of the background universe, i.e., the expansion history. We thus regard the DGP model as a ``dark energy model''. The main difference between DE and MG models usually comes from the aspect of growth of structure (see, e.g., Refs. \cite{11,22}), but we do not discuss this aspect in this paper. Note also that when we count the number of parameters of dark energy models, $k$, we include the dimensionless Hubble constant $h$.

The constraint results for these dark energy models using the current observational data are given in Table \ref{table1}. The results of the model comparison using the information criteria are summarized in Table \ref{table2}.

\begin{table*}\tiny
\caption{Fit results for the dark energy models by using the current data.}
\label{table1}
\small
\setlength\tabcolsep{10.8pt}
\renewcommand{\arraystretch}{1.5}
\centering
\begin{tabular}{cccccccccccc}
\\
\hline
Model  & \multicolumn{3}{c}{{ Parameter}}  \\ \hline

$\Lambda$CDM       & $h=0.667^{+0.006}_{-0.005}$
                   & $\Omega_{\rm{m}}=0.324^{+0.007}_{-0.008}$

                   \\

DGP                & $h=0.601^{+0.004}_{-0.006}$
                   & $\Omega_{\rm{m}}=0.367^{+0.004}_{-0.006}$

                   \\
NADE               & $h=0.629^{+0.004}_{-0.004}$
                   & $n=2.455^{+0.034}_{-0.033}$

                   \\
$w$CDM             & $h=0.662^{+0.008}_{-0.007}$
                   & $\Omega_{\rm{m}}=0.326^{+0.009}_{-0.008}$
                   & $w=-0.964^{+0.030}_{-0.036}$

                   \\
HDE                & $h=0.655^{+0.007}_{-0.007}$
                   & $\Omega_{\rm{m}}=0.326^{+0.009}_{-0.008}$
                   & $c=0.733^{+0.040}_{-0.039}$

                   \\
RDE                & $h=0.664^{+0.005}_{-0.005}$
                   & $\Omega_{\rm{m}}=0.350^{+0.007}_{-0.006}$
                   & $\gamma=0.325^{+0.009}_{-0.010}$

                   \\
$\alpha$DE         & $h=0.663^{+0.007}_{-0.008}$
                   & $\Omega_{{\rm m}}=0.326^{+0.008}_{-0.008}$
                   & $\alpha=0.106^{+0.140}_{-0.111}$

                      \\
GCG                & $h=0.663^{+0.008}_{-0.007}$
                   & $A_{\rm{s}}=0.695^{+0.024}_{-0.023}$
                   & $\beta=-0.03^{+0.067}_{-0.057}$

                      \\
CPL                & $h=0.663^{+0.007}_{-0.008}$
                   & $\Omega_{\rm{m}}=0.326^{+0.009}_{-0.007}$
                   & $w_{0}=-0.969^{+0.098}_{-0.094}$
                   & $w_{a}=0.007^{+0.366}_{-0.431}$

                      \\
NGCG               & $h=0.662^{+0.015}_{-0.014}$
                   & $\Omega_{\rm{de}}=0.673^{+0.008}_{-0.007}$
                   & $w=-0.969^{+0.031}_{-0.041}$
                   & $\eta=1.004^{+0.013}_{-0.010}$

                      \\

\hline
\end{tabular}
\end{table*}

\begin{table*}\tiny
\caption{Summary of the information criteria results.}
\label{table2}
\small
\setlength\tabcolsep{40.5pt}
\renewcommand{\arraystretch}{1.5}
\centering
\begin{tabular}{cccccccccccc}
\\
\hline
Model &$\chi^{2}_{{\rm min}}$  &$\Delta$AIC& $\Delta$BIC \\ \hline

$\Lambda$CDM
                   &$ 699.375$
                   &$ 0$
                   &$ 0$
                   \\
GCG
                   &$ 698.381$
                   &$ 1.006$
                   &$ 5.623$
                   \\

$w$CDM
                   &$ 698.524$
                   &$ 1.149$
                   &$ 5.766$
                   \\
$\alpha$DE
                   &$698.574$
                   &$ 1.199$
                   &$ 5.816$
                   \\

HDE
                   &$ 704.022$
                   &$ 6.647$
                   &$ 11.264$
                   \\

NGCG
                   &$698.331$
                   &$2.956$
                   &$12.191$
                   \\

CPL
                   &$ 698.543$
                   &$ 3.199$
                   &$ 12.401$
                   \\

NADE
                   &$ 750.229$
                   &$ 50.854$
                   &$ 50.854$
                   \\
DGP
                   &$ 786.326$
                   &$ 86.951$
                   &$ 86.951$
                   \\
RDE
                   &$987.752$
                   &$290.337$
                   &$294.994$
                   \\

\hline
\end{tabular}
\end{table*}

\subsection{Cosmological constant model}

The cosmological constant $\Lambda$ has nowadays become the most promising candidate for dark energy responsible for the current acceleration of the universe, because it can explain the various observations quite well, although it has been suffering the severe theoretical puzzles. The cosmological model with $\Lambda$ and CDM is called the $\Lambda$CDM model. Since the EoS of the vacuum energy (or $\Lambda$) is $w=-1$, we have
\begin{equation}
E(z)=\left[\Omega_{\rm{m}}(1+z)^{3}+\Omega_{\rm{r}}(1+z)^{4}+(1-\Omega_{\rm{m}}-\Omega_{\rm{r}})\right]^{1/2}.
\end{equation}

By using the observational data described in the above section, we can obtain the best-fit values of parameters and the corresponding $\chi^{2}_{\rm{min}}$,
\begin{equation}
\Omega_{\rm{m}}=0.324,  \quad h=0.667,  \quad \chi^{2}_{\rm{min}}=699.375.
\end{equation}
We also show the 1--2$\sigma$ posterior distribution contours in the $\Omega_{\rm{m}}$--$h$ plane for the $\Lambda$CDM model in Fig.~\ref{figg1}.

Among the models discussed in this paper, the $\Lambda$CDM model has the lowest AIC and BIC values, which shows that this model is still the most favored cosmological model by current data nowadays. We thus choose the $\Lambda$CDM model as the reference model in the model comparison, i.e., the values of $\Delta {\rm AIC}$ and $\Delta {\rm BIC}$ of other models are measured relative to this model.

\begin{figure*}
\includegraphics[width=8cm]{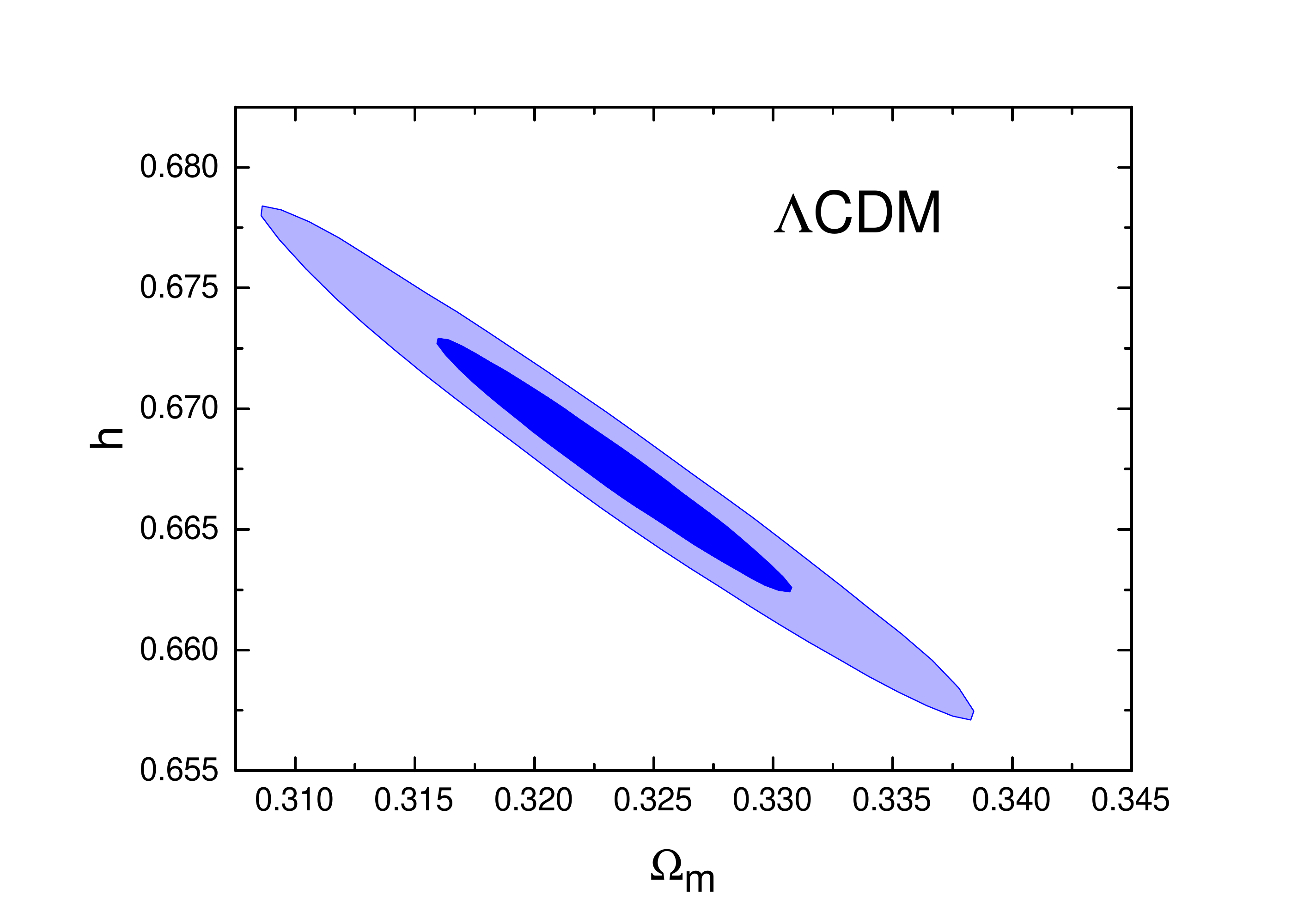}\\
\caption{\label{figg1}The cosmological constant model: 68.3\% and 95.4\% confidence level contours in the $\Omega_{\rm{m}}$--$h$ plane.}
\end{figure*}

\subsection{Dark energy models with equation of state parameterized}

In this class, we consider two models: the constant $w$ parametrization ($w$CDM) model and the Chevallier-Polarski-Linder (CPL) parametrization model.

\subsubsection{Constant $w$ parametrization}

In this model, one assumes that the EoS of dark energy is $w={\rm constant}$. This is the simplest case for a dynamical dark energy. It is hard to believe that this model would correspond to the real physical situation, but it can describe dynamical dark energy in a simply way. This model is also called the $w$CDM model. In this model, we have
\begin{equation}
\begin{aligned}
E(z)^{2}&=\Omega_{\rm{m}}(1+z)^{3}+\Omega_{\rm{r}}(1+z)^{4}\\
&+(1-\Omega_{\rm{m}}-\Omega_{\rm{r}})(1+z)^{3(1+w)},
\end{aligned}
\end{equation}

According to the observations, the best-fit parameters and the corresponding $\chi^{2}_{\rm{min}}$ are
\begin{equation}
\Omega_{\rm{m}}=0.326, \; h=0.662, \; w=-0.964, \; \chi^{2}_{\rm{min}}=698.524.
\end{equation}

The 1--2$\sigma$ posterior possibility contours in the $\Omega_{\rm{m}}$--$w$ and $\Omega_{\rm{m}}$--$h$ planes for the $w$CDM model are plotted in Fig.~\ref{fig2}. We find that the constraint result of $w$ is consistent with the cosmological constant at about the 1$\sigma$ level. Compared to the $\Lambda$CDM model, this model yields a lower $\chi^2$, due to the fact that it has one more parameter, and this has been punished by the information criteria, $\Delta {\rm AIC}=1.149$ and $\Delta {\rm BIC}=5.766$.

\begin{figure*}
\includegraphics[width=8cm]{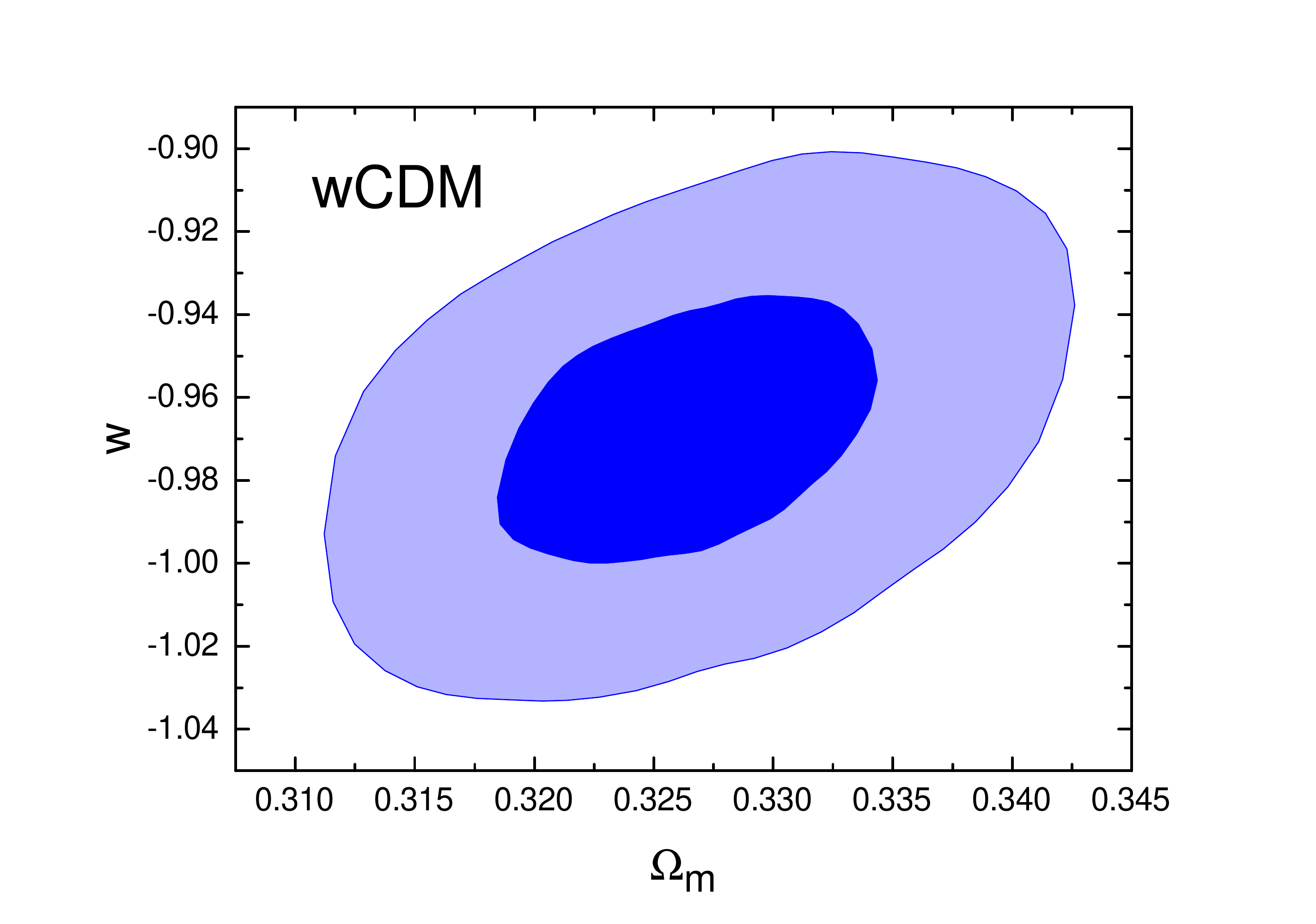}
\includegraphics[width=8cm]{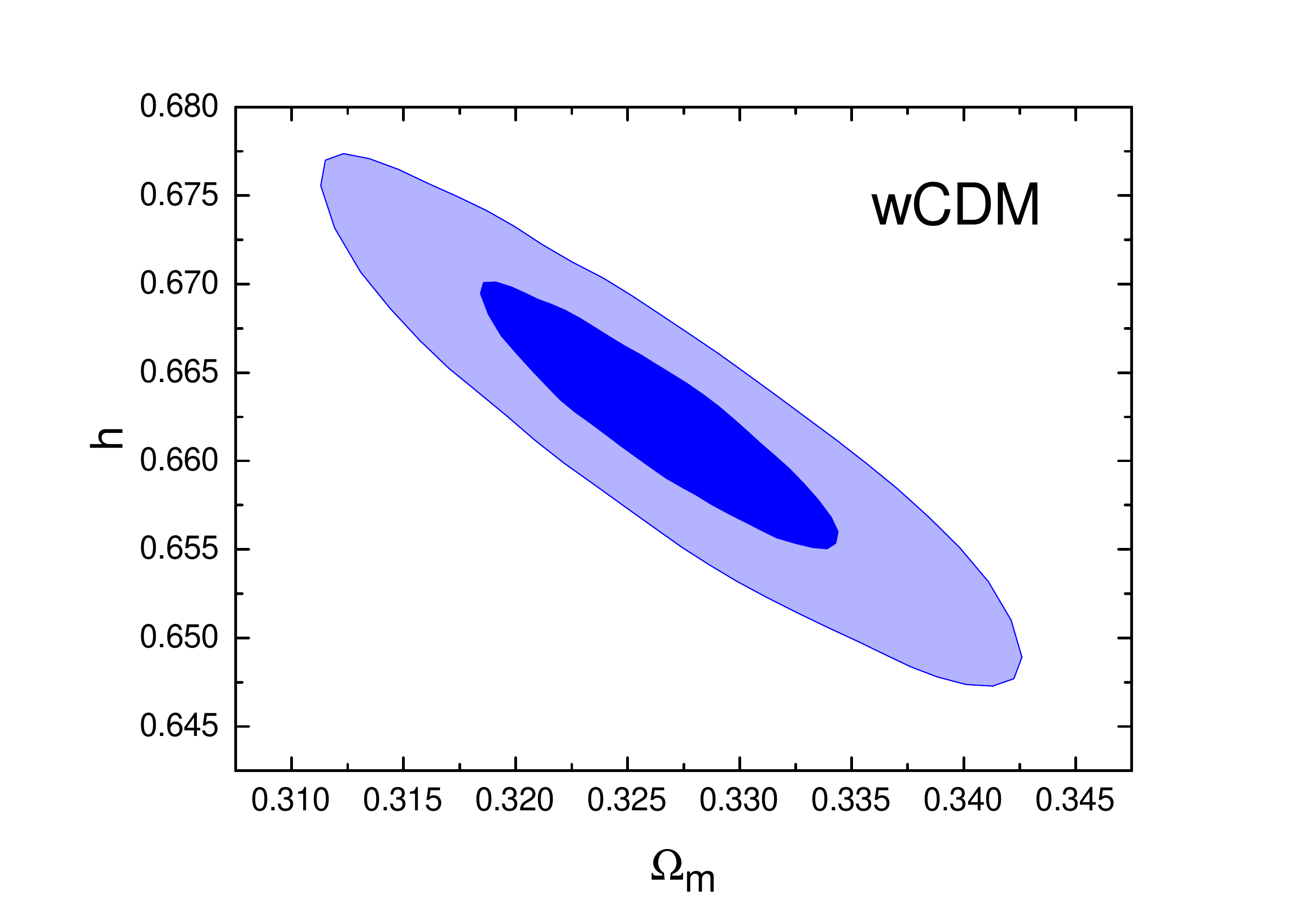}\\
\caption{\label{fig2}The constant $w$ model: 68.3\% and 95.4\% confidence level contours in the $\Omega_{\rm{m}}$--$w$ and $\Omega_{\rm{m}}$--$h$ planes.}
\end{figure*}

\subsubsection{Chevallier-Porlarski-Linder parametrization}

To probe the evolution of $w$ phenomenologically, the most widely used parametrization model is the CPL model~\cite{MD:2001,EV:2003}, sometimes called $w_0w_a$CDM model. For this model, the form of $w(z)$ is written as
\begin{equation}
w(z)=w_{\rm{0}}+w_{\rm{a}}\frac{z}{1+z},
\end{equation}
where $w_{0}$ and $w_{{\rm a}}$ are free parameters. This parametrization has some advantages such as high accuracy in reconstructing scalar field equation of state and has simple physical interpretation. Detailed description can be found in Ref.~\cite{EV:2003}. For this model, we have
\begin{equation}
\begin{aligned}
E(z)^{2}&=\Omega_{\rm{m}}(1+z)^{3}+\Omega_{\rm{r}}(1+z)^{4}\\
&+(1-\Omega_{\rm{m}}-\Omega_{\rm{r}})(1+z)^{3(1+w_{\rm{0}}+w_{\rm{a}})}\exp\left(-\frac{3w_{\rm{a}}z}{1+z}\right).
 \end{aligned}
\end{equation}

The joint observational constraints give the best-fit parameters and the corresponding $\chi^{2}_{\rm{min}}$:
\begin{equation}
\begin{aligned}
&\Omega_{\rm{m}}=0.326,   \quad w_{\rm{0}}=-0.969,   \quad  w_{\rm{a}}=0.007,\\
 &h=0.663,   \quad \chi^{2}_{\rm{min}}=698.543.
\end{aligned}
\end{equation}
The 1--2$\sigma$ likelihood contours for the CPL model in the $w_{\rm{0}}$--$w_{\rm{a}}$ and $\Omega_{\rm{m}}$--$h$ planes are shown in Fig.~\ref{fig3}.

We find that the constraint result of the CPL model is consistent with the $\Lambda$CDM model, i.e., the point of $\Lambda$CDM ($w_0=-1$ and $w_a=0$) still lies in the 1$\sigma$ region (on the edge of 1$\sigma$). The CPL model has two more parameters than $\Lambda$CDM, so that it yields a lower $\chi^2$, but the difference $\Delta \chi^2=-0.832$ is rather small. The AIC punishes the CPL model on the number of parameters, leading to $\Delta {\rm AIC}=3.199$, and furthermore the BIC punishes it on the number of data points, leading to $\Delta {\rm BIC}=12.401$.

\begin{figure*}
\includegraphics[width=8cm]{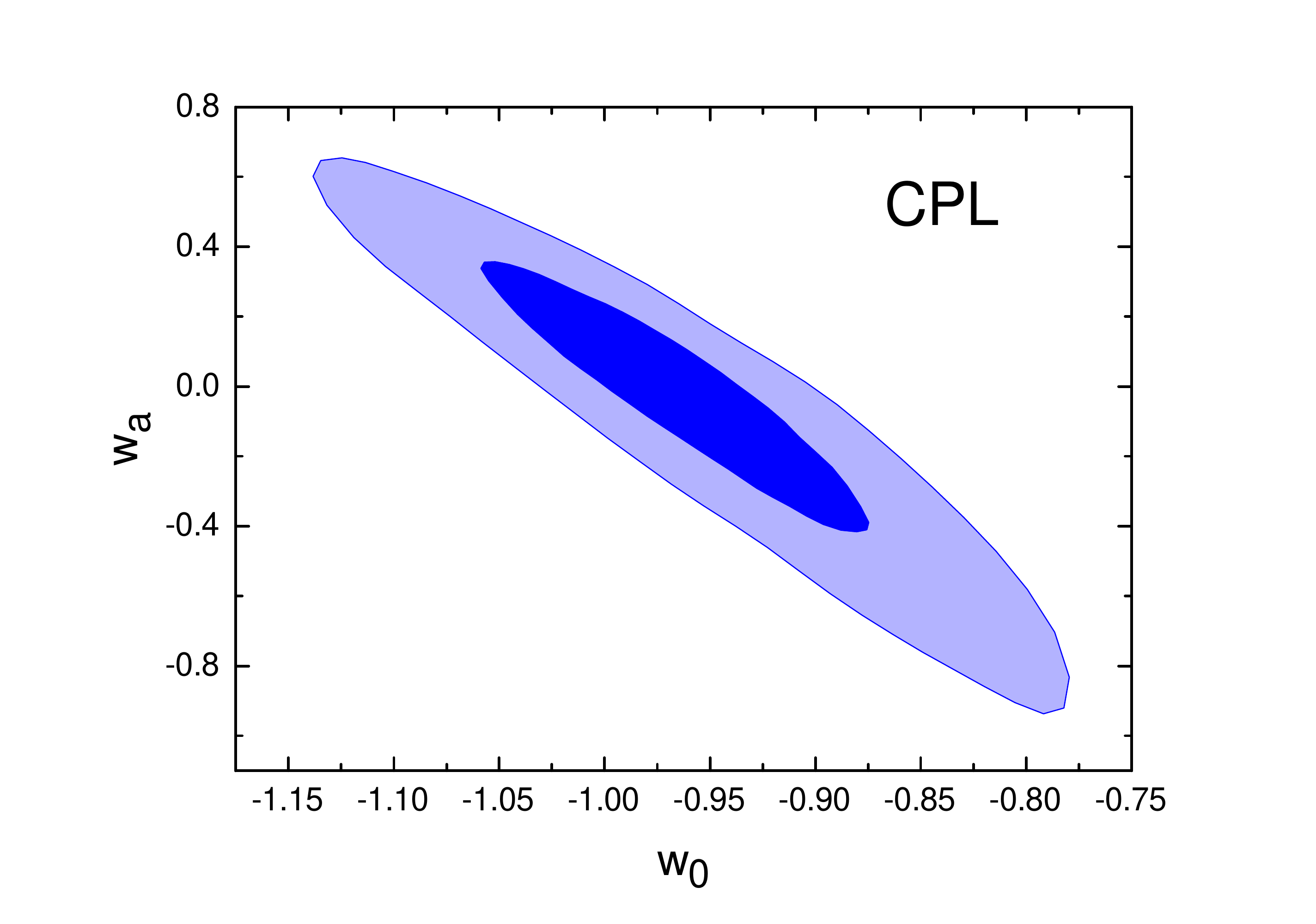}
\includegraphics[width=8cm]{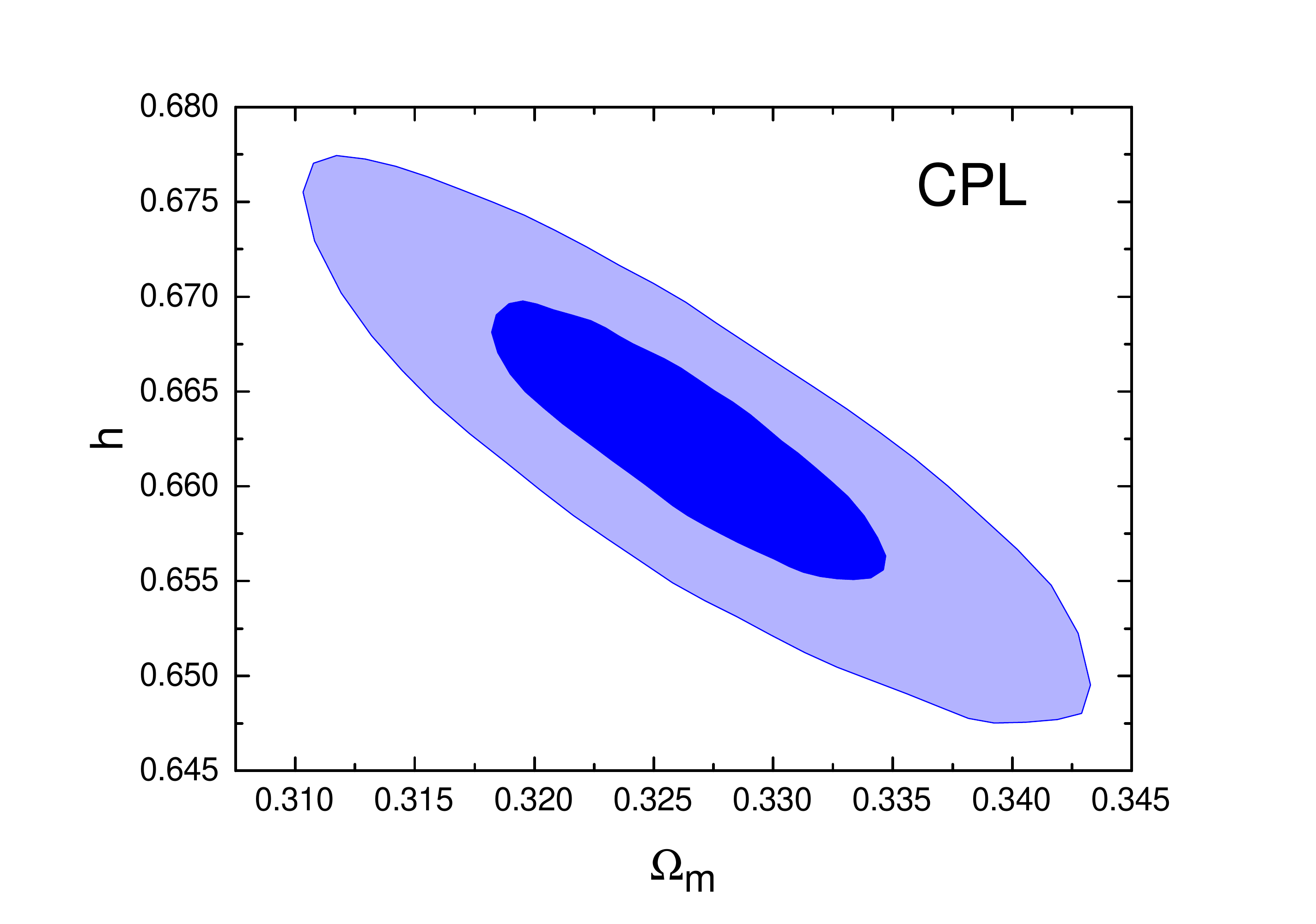}\\
\caption{\label{fig3}The Chevallier-Polarski-Linder model: 68.3\% and 95.4\% confidence level contours in the $w_{\rm{0}}$--$w_{\rm{a}}$ and $\Omega_{\rm{m}}$--$h$ planes.}
\end{figure*}

\subsection{Chaplygin gas models}

The Chaplygin gas model \cite{Kamenshchik:2001cp}, which is commonly viewed as arising from the $d$-brane theory, can describe the cosmic acceleration, and it provides a unification scheme for vacuum energy and cold dark matter. The original Chaplygin gas model has been excluded by observations \cite{c:2003}, thus here we only consider the generalized Chaplygin gas (GCG) model \cite{MCB:2002} and the new generalized Chaplygin gas (NGCG) model \cite{Zhang:2004gc}. These models can be viewed as interacting dark energy models with the interaction term $Q\propto {\rho_{\rm de}\rho_{\rm c}\over \rho_{\rm de}+\rho_{\rm c}}$, where $\rho_{\rm de}$ and $\rho_{\rm c}$ are the energy densities of dark energy and cold dark matter~\cite{Li:2013bya}.

\subsubsection{Generalized Chaplygin gas model}

The GCG has an exotic equation of state,
\begin{equation}
p_{\rm{gcg}}=-\frac{A}{\rho^{\beta}_{\rm{gcg}}},
\end{equation}
where A is a positive constant and $\beta$ is a free parameter. Thus, the energy density of GCG can be derived,
\begin{equation}
\rho_{\rm{gcg}}(a)=\rho_{\rm{gcg}0}\left(A_{\rm{s}}+\frac{1-A_{\rm{s}}}{a^{3(1+\beta)}}\right)^{\frac{1}{1+\beta}},
\end{equation}
where $A_{\rm{s}}\equiv A/\rho^{1+\beta}_{\rm{gcg}0}$. It is obvious that the GCG behaves as a dust-like matter at the early times and behaves like a cosmological constant at the late stage. In this model, we have
\begin{equation}
\begin{aligned}
E(z)^{2}&=\Omega_{\rm{b}}(1+z)^{3}+\Omega_{\rm{r}}(1+z)^{4}\\
&+(1-\Omega_{\rm{b}}-\Omega_{\rm{r}})\left(A_{\rm{s}}+(1-A_{\rm{s}})(1+z)^{3(1+\beta)}\right)^{1\over 1+\beta}.
\end{aligned}
\end{equation}
It should be noted that the cosmological constant model is recovered for $\beta=0$ and $\Omega_{\rm{m}}=1-\Omega_{\rm{r}}-A_{\rm{s}}(1-\Omega_{\rm{r}}-\Omega_{\rm{b}})$. 

Through the joint data analysis, we get the  best-fit parameters and the corresponding $\chi^{2}_{\rm{min}}$:
\begin{equation}
 A_{\rm{s}}=0.695,\, \beta=-0.03, \,  h=0.663,\,\chi^{2}_{\rm{min}}=698.381.
\end{equation}
We show the likelihood contours for the GCG model in the $A_{\rm{s}}$--$\beta$ and $A_{\rm{s}}$--$h$ planes in Fig.~\ref{fig4}. 

From the constraint results, we can see that the value of $\beta$ is close to zero, which implies that the $\Lambda$CDM limit of this model is favored. For the GCG model, we have $\Delta {\rm AIC}=1.006$ and $\Delta {\rm BIC}=5.623$.

\begin{figure*}
\includegraphics[width=8cm]{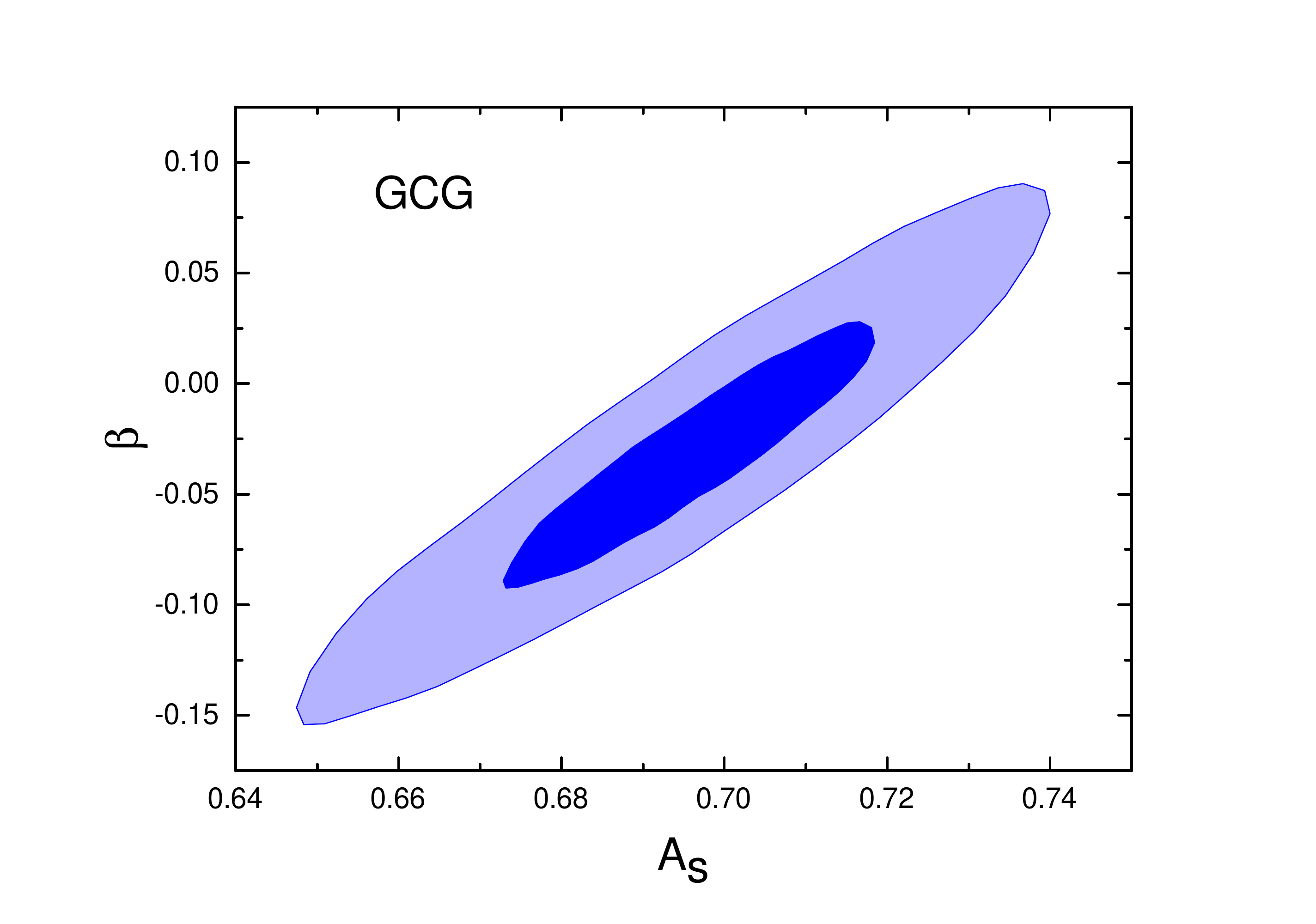}
\includegraphics[width=8cm]{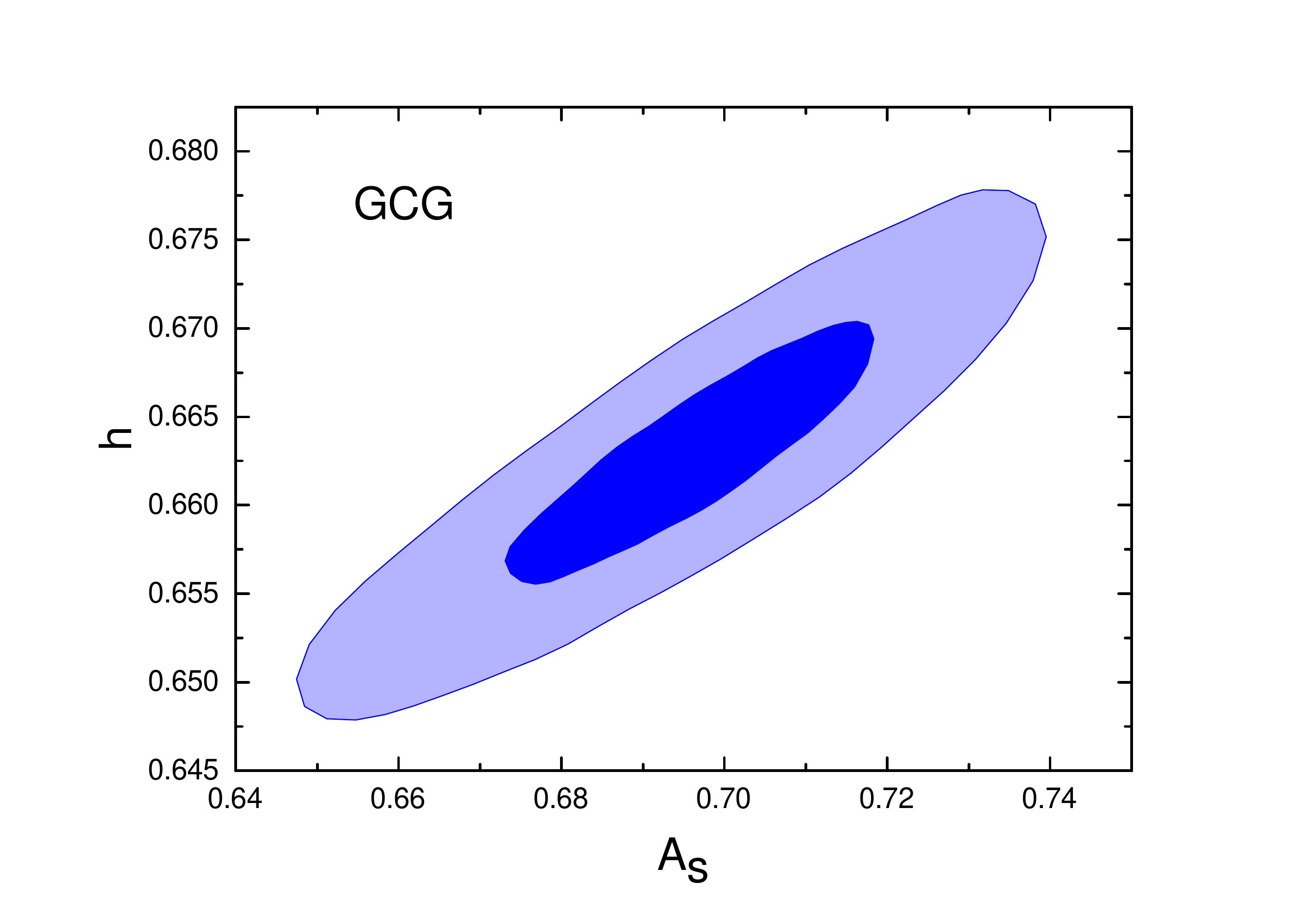}\\
\caption{\label{fig4}The generalized Chaplygin gas model: 68.3\% and 95.4\% confidence level contours in the $A_{\rm{s}}$--$\beta$ and $A_{\rm{s}}$--$h$ planes.}
\end{figure*}

\subsubsection{New generalized Chaplygin gas model}

The GCG model actually can be viewed as an interacting model of vacuum energy with cold dark matter. If one wishes to further extend the model, a natural idea is that the vacuum energy is replace with a dynamical dark energy. Thus, the NGCG model was proposed \cite{Zhang:2004gc}, in which the dark energy with constant $w$ interacts with cold dark matter through the interaction term $Q=-3\beta w H {\rho_{\rm de}\rho_{\rm c}\over \rho_{\rm de}+\rho_{\rm c}}$. That is to say, this model is actually a type of interacting $w$CDM model. Such an interacting dark energy model is a large-scale stable model, naturally avoiding the usual super-horizon instability problem existing in the interacting dark energy models \cite{Li:2013bya}. (The large-scale instability problem in the interacting dark energy models has been systematically solved by establishing a parameterized post-Friedmann framework for interacting dark energy \cite{Li:2014eha,Li:2014cee,Li:2015vla}.) The model has recently been investigated in detail in Ref. \cite{Li:2013bya}.

The equation of state of the NGCG fluid \cite{Zhang:2004gc} is given by
\begin{equation}
p_{\rm{ngcg}}=-\frac{\tilde{A}(a)}{\rho^{\beta}_{\rm{ngcg}}},
\end{equation}
where $\tilde{A}(a)$ is a function of the scale factor $a$ and $\beta$ is a free parameter. The energy density of the NGCG can be expressed as 
\begin{equation}
\rho_{\rm{ngcg}}=\left[Aa^{-3(1+w)(1+\beta)}+Ba^{-3(1+\beta)}\right]^{\frac{1}{1+\beta}},\label{ngcg1}
\end{equation}
where $A$ and $B$ are positive constant. 
The form of the function $\tilde{A}(a)$ can be determined to be
\begin{equation}
\tilde{A}(a)=-wAa^{-3(1+w)(1+\beta)}.
\end{equation}
Considering a universe with NGCG, baryon, and radiation, we can get
\begin{equation}
\begin{aligned}
E(z)^{2}&=\Omega_{\rm{b}}(1+z)^{3}+\Omega_{\rm{r}}(1+z)^{4}+(1-\Omega_{\rm{b}}-\Omega_{\rm{r}})(1+z)^{3}\\
&\left[1-\frac{\Omega_{\rm{de}}}{1-\Omega_{\rm{b}}-\Omega_{\rm{r}}}\left(1-(1+z)^{3w(1+\beta)}\right)\right]^{1\over 1+\beta}.\label{ngcg}
\end{aligned}
\end{equation}

The joint observational constraints give the best-fit parameters and the corresponding $\chi^{2}_{\rm{min}}$:
\begin{equation}
\begin{aligned}
&  \Omega_{\rm{de}}=0.673, \,    w=-0.969,\, \beta=0.004,  \\
& h=0.662, \quad  \chi^{2}_{\rm{min}}=698.331.
\end{aligned}
\end{equation}
We show the likelihood contours for the NGCG model in the $w$--$\eta$ and $\Omega_{{\rm de}}$--$h$ planes in Fig.~\ref{fig5}, where the parameter $\eta$ is defined as $\eta=1+\beta$ in \cite{Zhang:2004gc}.

The NGCG will reduce to GCG when $w=-1$, reduce to $w$CDM when $\eta=1$, and reduce to $\Lambda$CDM when $w=-1$ and $\eta=1$. From Fig.~\ref{fig5}, we see that the constraint results are consistent with GCG and $w$CDM within 1$\sigma$ range, and consistent with $\Lambda$CDM on the edge of 1$\sigma$ region. Though with more parameters, the NGCG model only yields a little bit lower $\chi^{2}_{\rm{min}}$ than the above sub-models, which is punished by the information criteria. For the NGCG model, we have $\Delta {\rm AIC}=2.956$ and $\Delta {\rm BIC}=12.191$.

\begin{figure*}
\includegraphics[width=8cm]{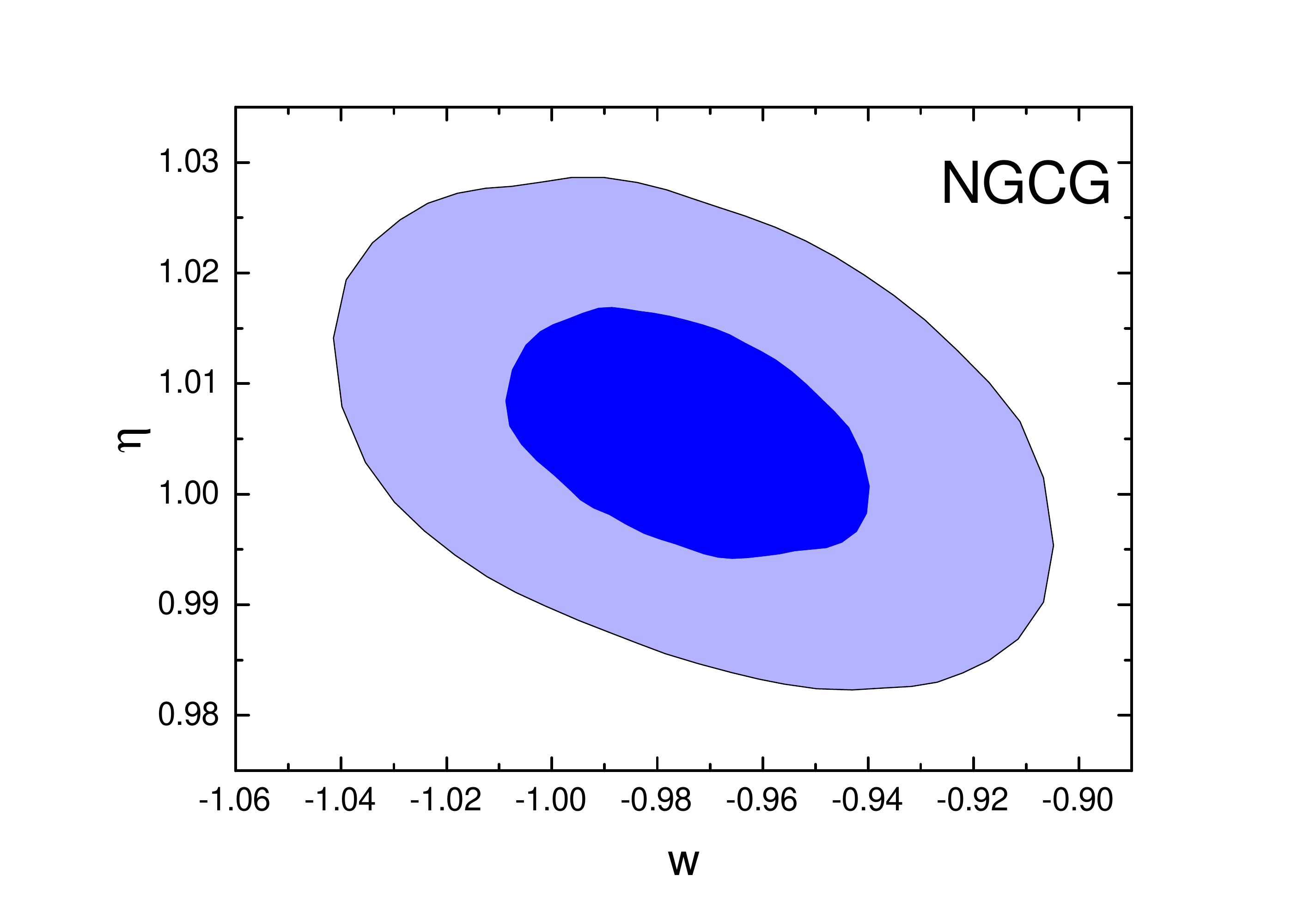}
\includegraphics[width=8cm]{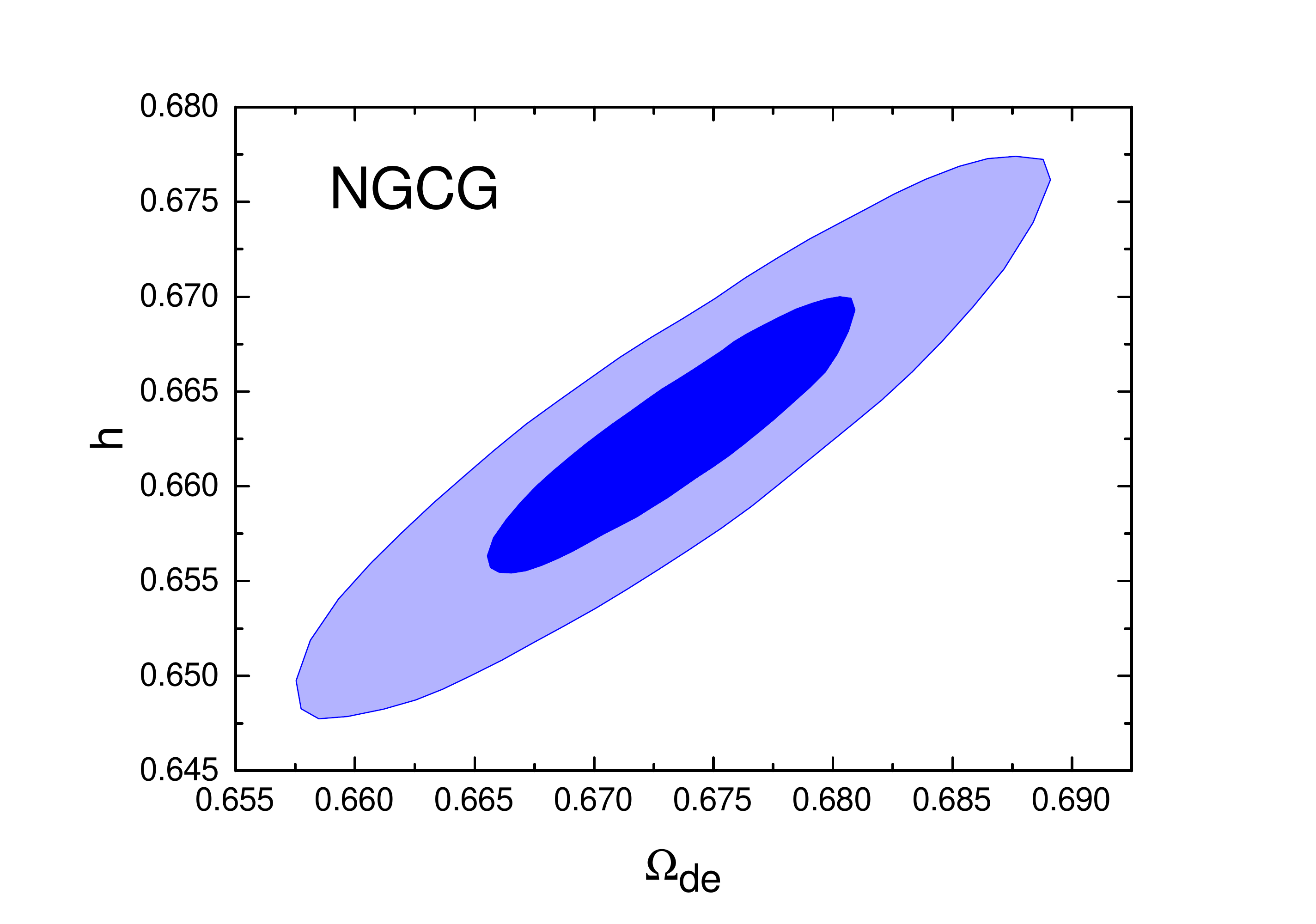}\\
\caption{\label{fig5}The new generalized Chaplygin gas model: 68.3\% and 95.4\% confidence level contours in the $w$--$\eta$ and $\Omega_{\rm{de}}$--$h$ planes.}
\end{figure*}

\subsection{Holographic dark energy models}

Within the framework of quantum field theory, the evaluated vacuum energy density will diverge; even though a reasonable ultraviolet (UV) cutoff is taken, the theoretical value of the vacuum energy density will still be larger than its observational value by several tens orders of magnitude. The root of this difficulty comes from the fact that a full theory of quantum gravity is absent. The holographic dark energy model was proposed under such circumstances, in which the effects of gravity is taken into account in the effective quantum field theory through the consideration of the holographic principle. When the gravity is considered, the number of degrees of freedom in a spatial region should be limited due to the fact that too many degrees of freedom would lead to the formation of a black hole \cite{A.G:1999wv}, which leads to the holographic dark energy model with the density of dark energy given by
\begin{equation}
\rho_{\rm{de}}\propto M^{2}_{\rm{pl}}L^{-2},
\end{equation}
where $L$ is the infrared (IR) cutoff length scale in the effective quantum field theory. Thus, in this sense, the UV problem of the calculation of vacuum energy density is converted to an IR problem. Different choices of the IR cutoff $L$ lead to different holographic dark energy models. In this paper, we consider three popular models in this setting: the HDE model \cite{M.Li:2004wv}, the NADE model~\cite{HWRG:2008}, and the RDE model~\cite{CGFQ:2009}.

\subsubsection{Holographic dark energy model}

The HDE model \cite{M.Li:2004wv} is defined by choosing the event horizon size of the universe as the IR cutoff in the holographic setting. The energy density of HDE is thus given by
\begin{equation}
\rho_{{\rm de}}=3c^{2}M^{2}_{\rm{pl}}R_{\rm{h}}^{-2},\label{HDE1}
\end{equation}
where $c$ is a dimensionless parameter which plays an important role in determining properties of the holographic dark energy and $R_{{\rm h}}$ is the future event horizon, defined as
\begin{equation}
R_{\rm{h}}(t)=ar_{\rm{max}}(t)=a(t)\int_t^\infty\frac{dt'}{a(t')}.\label{HDER}
\end{equation}
The evolution of the HDE is governed by the following differential equations,
\begin{equation}
\frac{1}{E(z)}\frac{dE(z)}{dz}=-\frac{\Omega_{\rm{de}}(z)}{1+z}\left(\frac{1}{2}+\frac{\sqrt{\Omega_{\rm{de}}(z)}}{c}-\frac{\Omega_{\rm{r}}(z)+3}{2\Omega_{\rm{de}}(z)}\right),\label{Ez}
\end{equation}
\begin{equation}
\begin{aligned}
\frac{d\Omega_{\rm{de}}(z)}{dz}&=-\frac{2\Omega_{\rm{de}}(z)(1-\Omega_{\rm{de}}(z))}{1+z} \\
&\left(\frac{1}{2}+\frac{\sqrt{\Omega_{\rm{de}}(z)}}{c}+\frac{\Omega_{\rm{r}}(z)}{2(1-\Omega_{\rm{de}}(z))}\right),\label{Ode}
\end{aligned}
\end{equation}
where the fractional density of radiation is defined as $\Omega_{\rm{r}}(z)=\Omega_{\rm{r}}(1+z)^4/E(z)^2$. 

For this model, from the joint observational data analysis, we get the best-fit parameters and the corresponding $\chi^{2}_{\rm{min}}$:
\begin{equation}
\Omega_{\rm{m}}=0.326, \;   c=0.733,    \;   h=0.655,    \; \chi^{2}_{\rm{min}}=704.022.
\end{equation}
We plot the likelihood contours for the HDE model in the $\Omega_{\rm{m}}$--$c$ and $\Omega_{{\rm m}}$--$h$ planes in Fig.~\ref{fig6}. 

The HDE model does not involve $\Lambda$CDM as a sub-model. Though it has one more parameter, it still yields a larger $\chi^{2}_{\rm{min}}$ than $\Lambda$CDM, showing that facing the current accurate data the HDE model behaves explicitly worse than $\Lambda$CDM. For the HDE model, we have $\Delta {\rm AIC}=6.647$ and $\Delta {\rm BIC}=11.264$.

\begin{figure*}
\includegraphics[width=8cm]{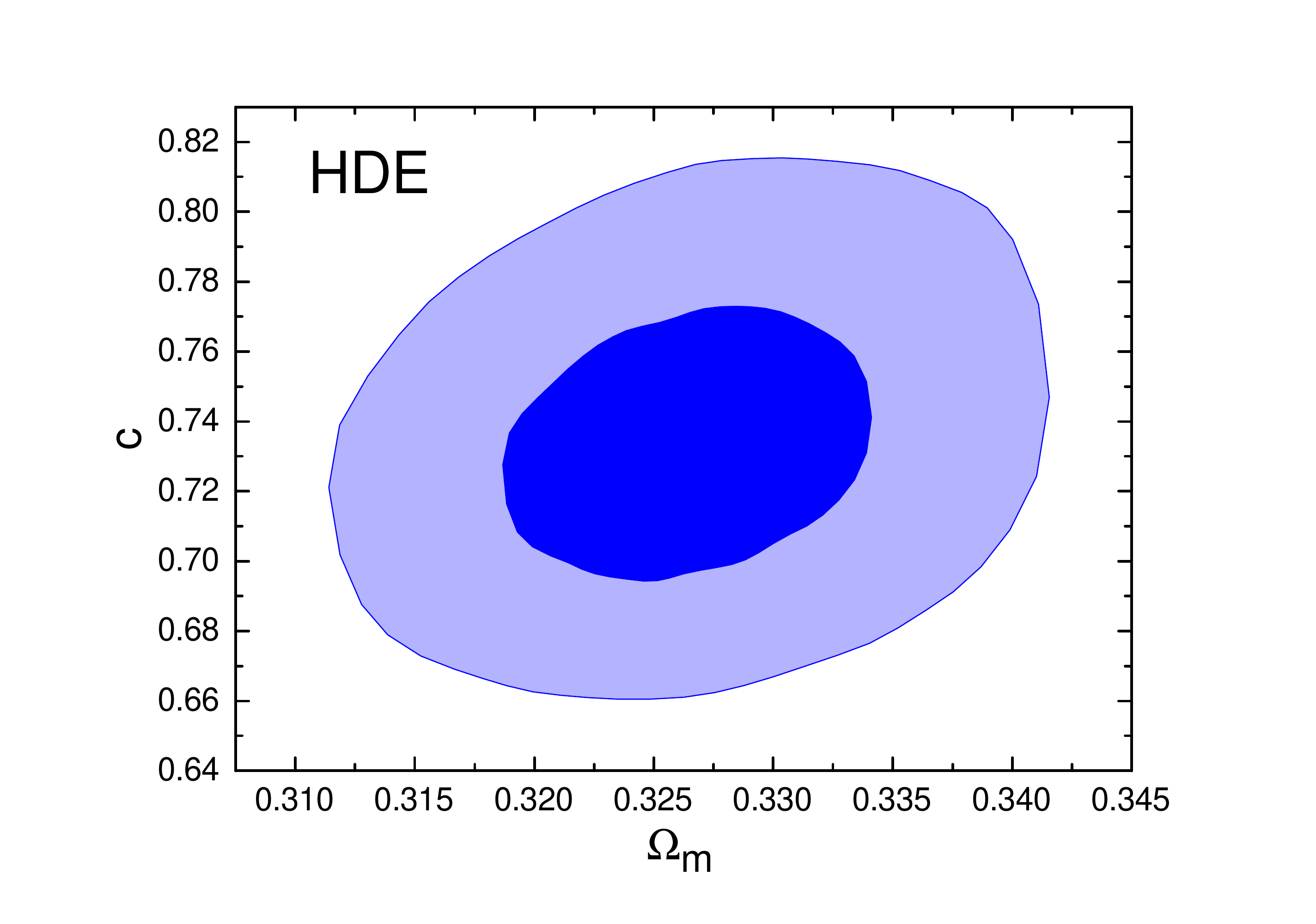}
\includegraphics[width=8cm]{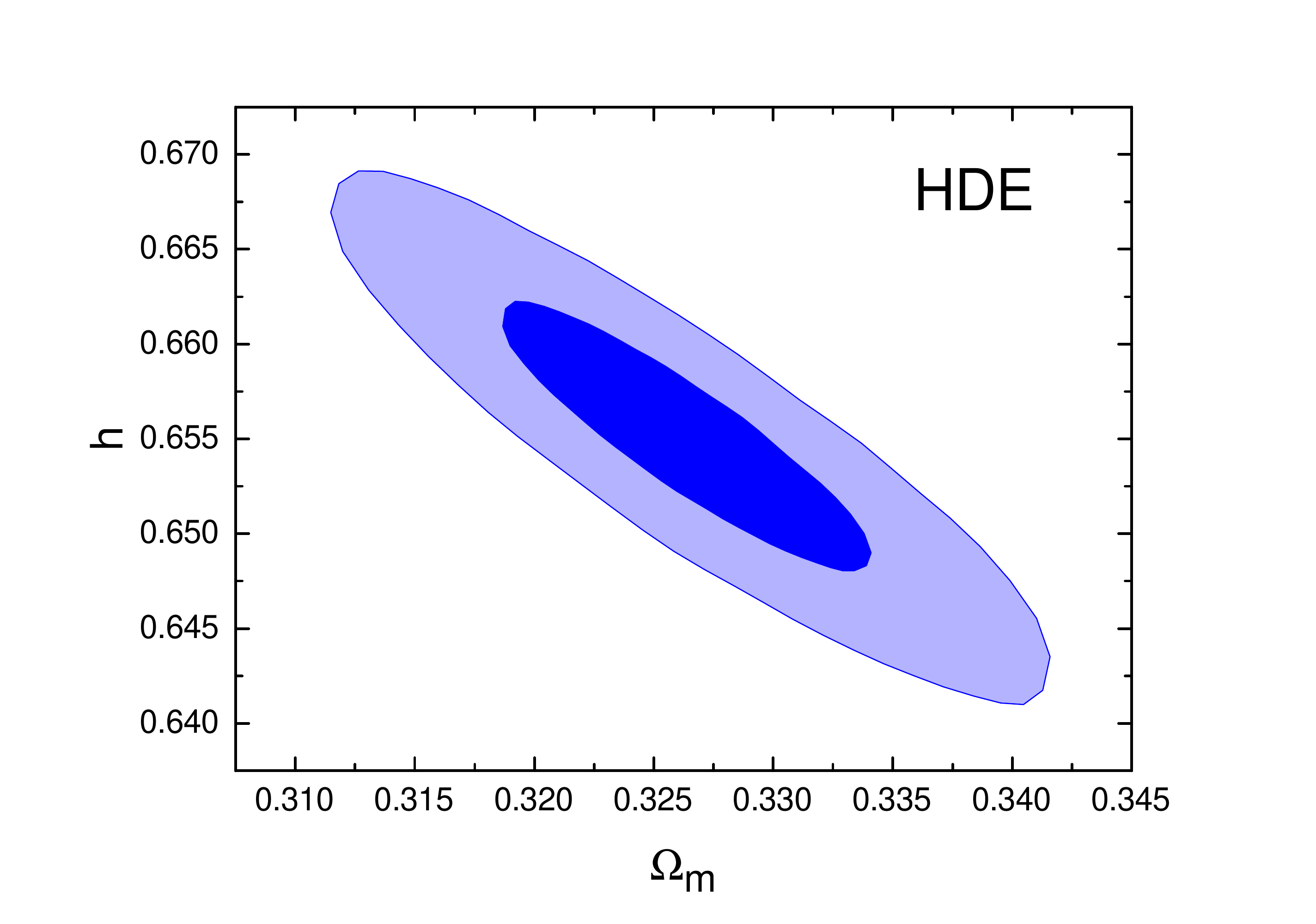}\\
\caption{\label{fig6}The holographic dark energy model: 68.3\% and 95.4\% confidence level contours in the $\Omega_{{\rm m}}$--$c$ and $\Omega_{{\rm m}}$--$h$ planes.}
\end{figure*}

\subsubsection{New agegraphic dark energy model}

The NADE model~\cite{HWRG:2008} chooses the conformal time of the universe $\tau$ as the IR cutoff in the holographic setting,
\begin{equation}
\tau=\int_0^t\frac{dt'}{a}=\int_0^a\frac{da'}{Ha'},
\end{equation}
so that the energy density of NADE is expressed as
\begin{equation}
\rho_{\rm{de}}=3n^{2}M^{2}_{\rm{pl}}\tau^{-2},
\end{equation}
where $n$ is a constant playing the same role as $c$ in the HDE model. In this model, the evolution of $\Omega_{{\rm de}}(z)$ is governed by the following differential equation:
\begin{equation}
\begin{aligned}
\frac{d\Omega_{\rm{de}}(z)}{dz}&=-\frac{2\Omega_{\rm{de}}(z)(1-\Omega_{\rm{de}}(z))}{1+z}\\
&\left(\frac{3}{2}-\frac{(1+z)\sqrt{\Omega_{\rm{de}}(z)}}{n}+\frac{\Omega_{\rm{r}}(1+z)}{2(\Omega_{\rm{m}}+\Omega_{\rm{r}}(1+z))}\right).\label{nOde}
\end{aligned}
\end{equation}
Then $E(z)$ can be derived,
\begin{equation}
E(z)=\left[\frac{\Omega_{{\rm m}}(1+z)^{3}+\Omega_{{\rm r}}(1+z)^{4}}{1-\Omega_{{\rm de}}(z)}\right]^{1/2}.
\end{equation}

The NADE model has the same number of parameters as $\Lambda$CDM. The only free parameter in NADE is the parameter $n$, and $\Omega_{\rm m}$ is actually a derived parameter in this model. This is because in this model one can use the initial condition $\Omega_{{\rm de}}(z_{{\rm ini}})=\frac{n^{2(1+z_{{\rm ini}})^{-2}}}{4}(1+\sqrt{F(z_{{\rm ini}})})^{2}$ at $z_{\rm{ini}}=2000$, with $F(z)\equiv\frac{\Omega_{{\rm r}}(1+z)}{\Omega_{{\rm m}}+\Omega_{{\rm r}}(1+z)}$, to solve Eq.~(\ref{nOde}). Once Eq.~(\ref{nOde}) is solved, one can then use $\Omega_{\rm{m}}=1-\Omega_{\rm{de}}(0)-\Omega_{\rm{r}}$ to get the value of $\Omega_{{\rm m}}$ (for detailed discussions, we refer the reader to Refs.~\cite{Li:2012xm,HWRG:200811,Zhang:2012pr}).

From the joint observational constraints, we get the best-fit parameters and the corresponding $\chi^{2}_{\rm{min}}$:
\begin{equation}
n=2.455,  \quad  h=0.629,  \quad \chi^{2}_{\rm{min}}=750.229.
\end{equation}
Based on the best-fit value of $n$, we can derive $\Omega_{\rm{m}}=0.336$. The likelihood contours for the NADE model in the $n$--$h$ plane is shown in Fig.~\ref{fig7}. 

We notice that the NADE model yields a large $\chi^{2}_{\rm{min}}$, much larger than that of $\Lambda$CDM. Since NADE and $\Lambda$CDM have the same number of parameters, the data-fitting capability can be directly compared through $\chi^{2}_{\rm{min}}$. For the NADE model, we have $\Delta {\rm AIC}=\Delta {\rm BIC}=50.854$. The constraint results show that, facing the precision cosmological observations, the NADE model cannot fit the current data well. 

\begin{figure*}
\includegraphics[width=8cm]{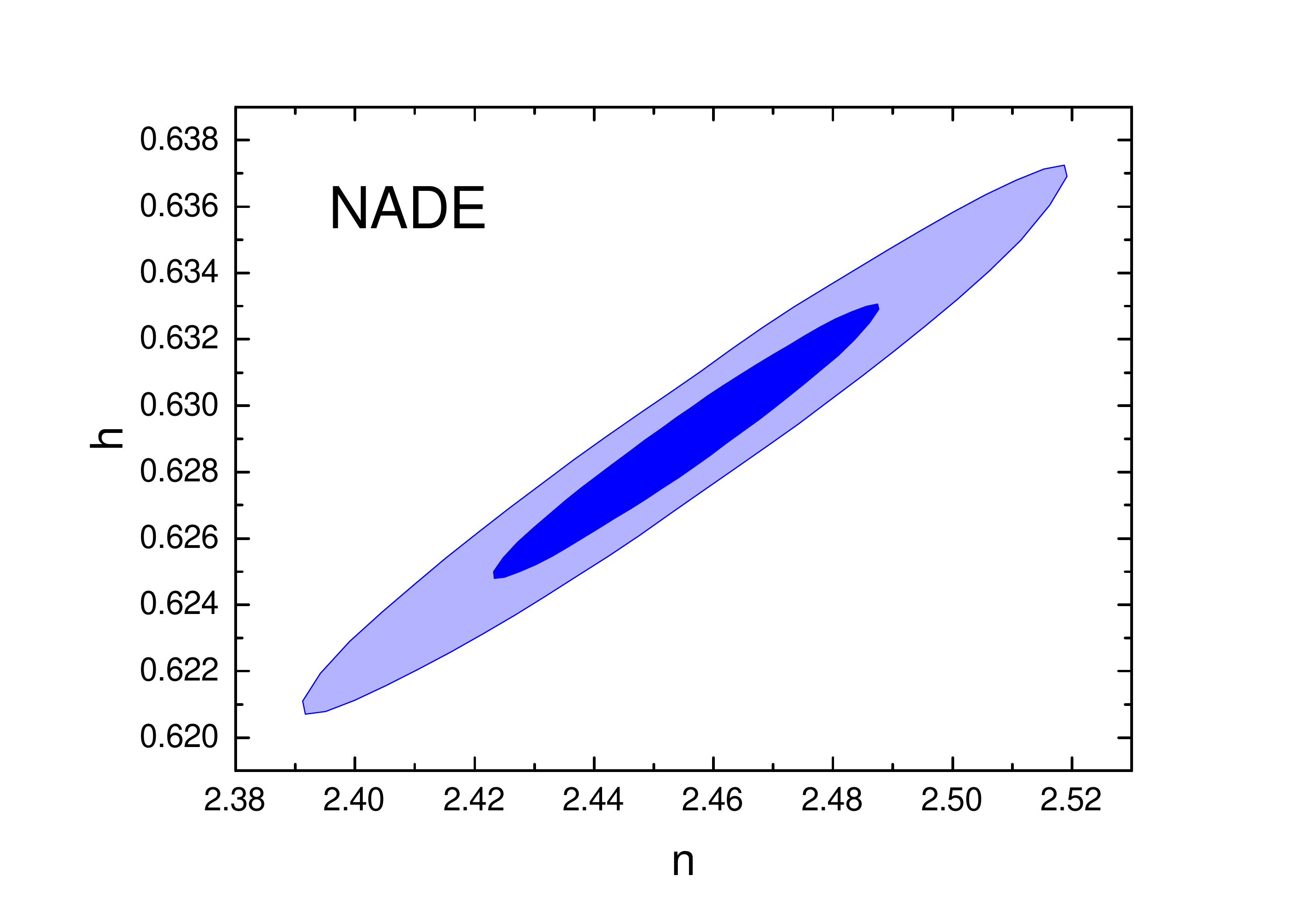}\\
\caption{\label{fig7}The new agegraphic dark energy model: 68.3\% and 95.4\% confidence level contours in the $n$--$h$ plane.}
\end{figure*}

\subsubsection{Ricci dark energy model}

The RDE model~\cite{CGFQ:2009} chooses the average radius of the Ricci scalar curvature as the IR cutoff length scale in the holographic setting (see also Refs. \cite{Cai:2008nk,Zhang:2009un}). In this model, the energy density of RDE can be expressed as
\begin{equation}
\rho_{\rm{de}}=3\gamma M^{2}_{\rm{pl}}(\dot{H}+2H^{2}),
\end{equation}
where $\gamma$ is a positive constant. The cosmological evolution in this model is determined by the following differential equation:
\begin{equation}
E^{2}=\Omega_{\rm{m}}e^{-3x}+\Omega_{\rm{r}}e^{-4x}+\gamma\left(\frac{1}{2}\frac{dE^{2}}{dx}+2E^{2}\right),
\end{equation}
where the $x=\ln a$. Solving this equation, we obtain
\begin{equation}
\begin{aligned}
E(z)^{2}&=\frac{2\Omega_{\rm{m}}}{2-\gamma}(1+z)^{3}+\Omega_{\rm{r}}(1+z)^{4}\\
&+\left(1-\Omega_{\rm{r}}-\frac{2\Omega_{\rm{m}}}{2-\gamma}\right)(1+z)^{(4-\frac{2}{\gamma})}.
\end{aligned}
\end{equation}

From the joint observational constraints, we get the best-fit parameters and the corresponding $\chi^{2}_{\rm{min}}$:
\begin{equation}
\Omega_{\rm{m}}=0.350,  \;  \gamma=0.325,   \; h=0.664,   \; \chi^{2}_{\rm{min}}=987.752.
\end{equation}
The likelihood contours for the RDE model in the $\Omega_{\rm m}$--$\gamma$ and $\Omega_{\rm m}$--$h$ planes are shown in Fig.~\ref{fig8}. 

We find that the RDE model yields a huge $\chi^{2}_{\rm{min}}$, much larger than those of other models considered in this model. For the RDE model, we have $\Delta {\rm AIC}=290.337$ and $\Delta {\rm BIC}=294.994$. The results of the observational constraints explicitly show that the RDE model has been excluded by the current observations.

\begin{figure*}
\includegraphics[width=8cm]{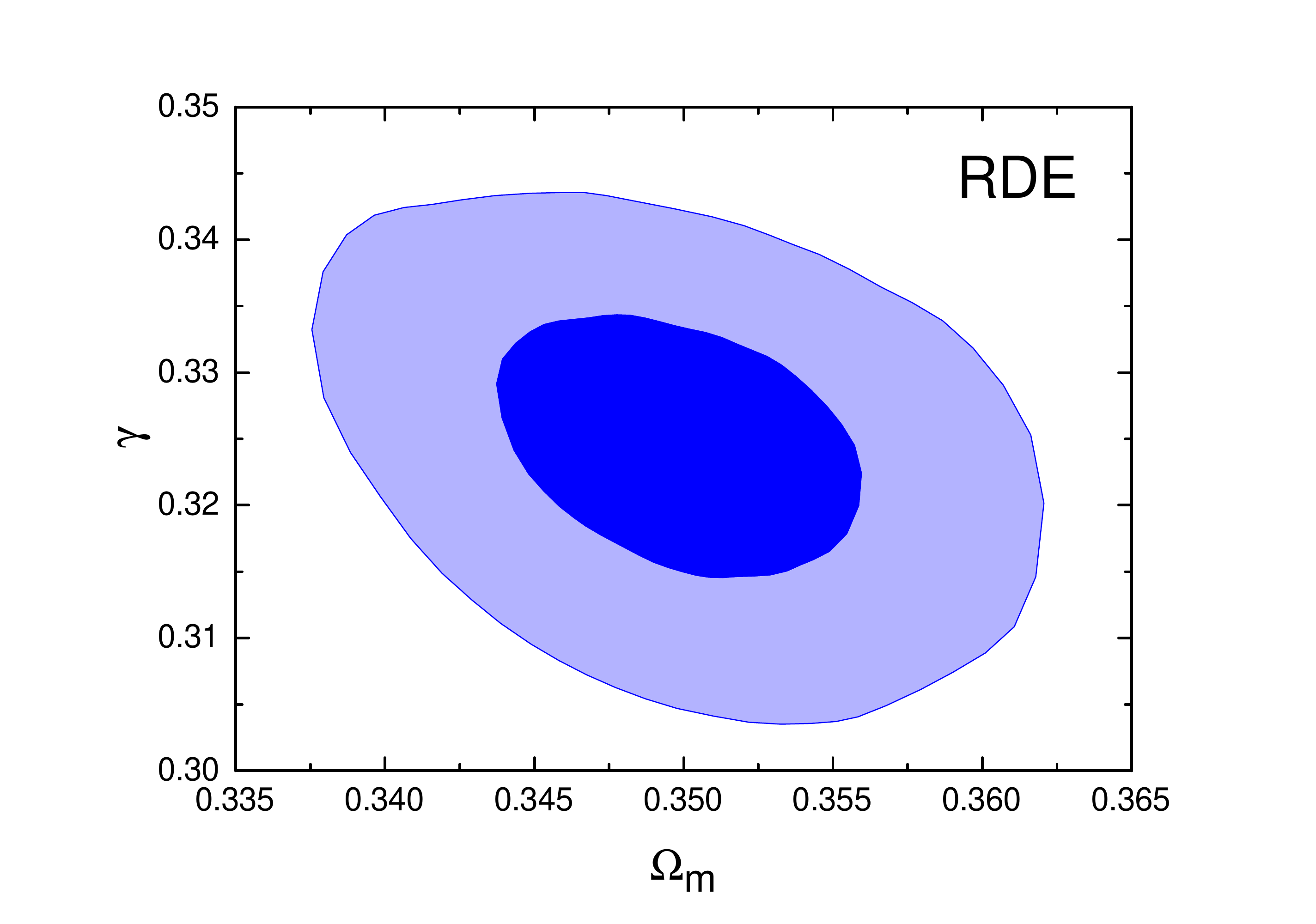}
\includegraphics[width=8cm]{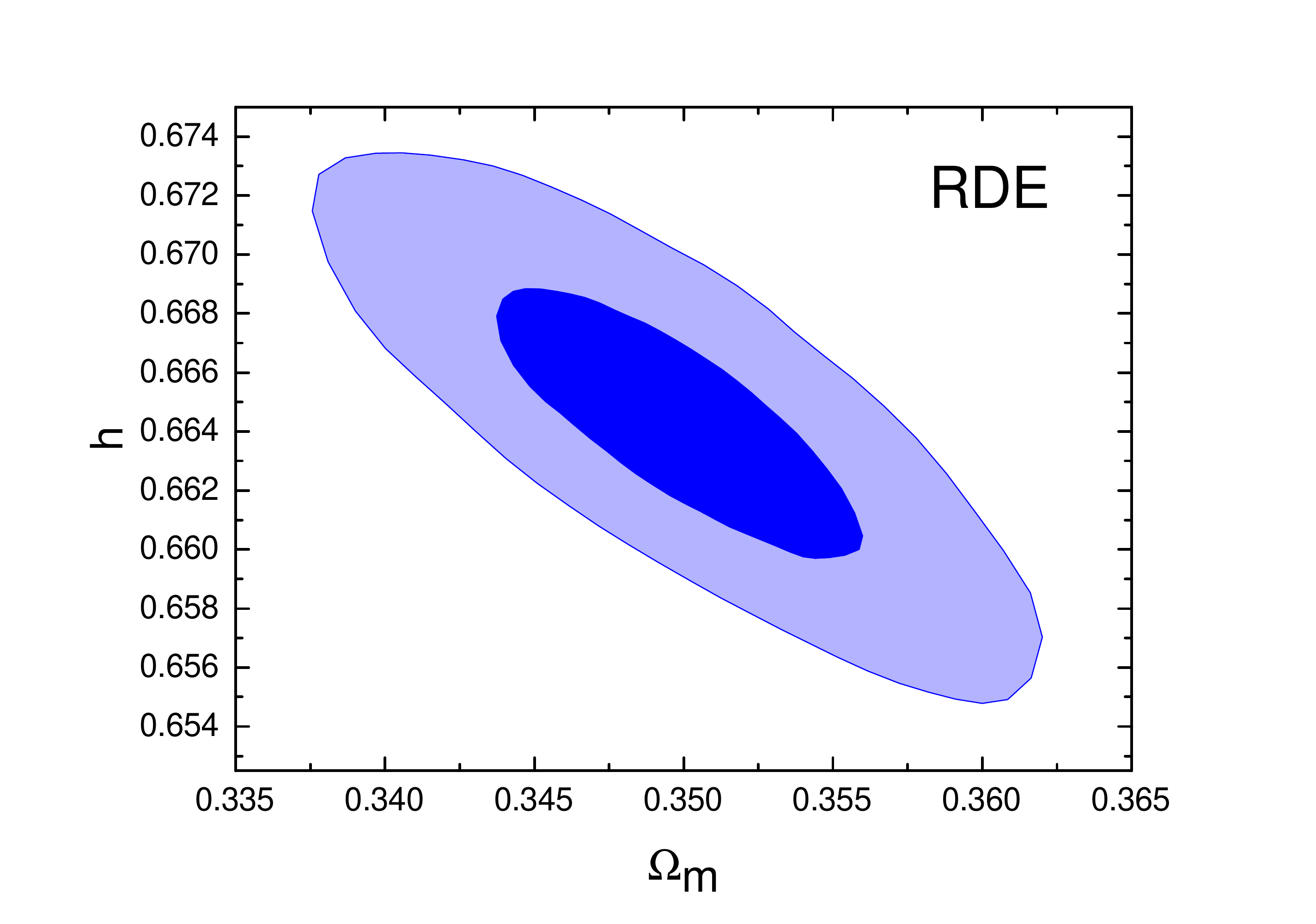}\\
\caption{\label{fig8}The Ricci dark energy model: 68.3\% and 95.4\% confidence level contours in the $\Omega_{\rm{m}}$--$\gamma$ and $\Omega_{\rm{m}}$--$h$ planes.}
\end{figure*}

\subsection{Dvali-Gabadadze-Porrati braneworld model and its phenomenological extension}

The DGP model~\cite{a:2000} is a well-known example of MG, in which a braneworld setting yields a self-acceleration of the universe without introducing dark energy. Inspired by the DGP model, a phenomenological model, called $\alpha$ dark energy model, was proposed in~\cite{c:2003}, which is much better than the DGP model in fitting the observational data.

\subsubsection{Dvali-Gabadadze-Porrati model }

In the DGP model \cite{a:2000}, the Friedmann equation is modified as
\begin{equation}
3M^{2}_{\rm{pl}}\left(H^{2}-\frac{H}{r_{\rm{c}}}\right)=\rho_{\rm{m}}(1+z)^{3}+\rho_{\rm{r}}(1+z)^{4},\label{DGP}
\end{equation}
where $r_c=[H_0(1-\Omega_{\rm m}-\Omega_{\rm r})]^{-1}$ is the crossover scale. Thus, $E(z)$ is given by
\begin{equation}
E(z)=\left[\sqrt{\Omega_{\rm{m}}(1+z)^{3}+\Omega_{\rm{r}}(1+z)^{4}+\Omega_{\rm{r_{c}}}}+\sqrt{\Omega_{\rm{r_{c}}}}\right],
\end{equation}
where $\Omega_{\rm{r_{c}}}=(1-\Omega_{\rm{m}}-\Omega_{\rm{r}})^{2}/4$. 

From the joint observational constraints, we get the best-fit parameters and the corresponding $\chi^{2}_{\rm{min}}$:\begin{equation}
\Omega_{\rm{m}}=0.367, \quad   h=0.601, \quad  \chi^{2}_{\rm{min}}=786.326.
\end{equation}
The likelihood contours for the DGP model in the $\Omega_{\rm{m}}$--$h$ is shown in Fig.~\ref{fig9}. 

The DGP model has the same number of parameters as $\Lambda$CDM. Compared to $\Lambda$CDM, the DGP model yields a much larger $\chi^{2}_{\rm{min}}$, indicating that the DGP model cannot fit the actual observations well. For the DGP model, we have $\Delta {\rm AIC}=\Delta {\rm BIC}=86.951$.

\begin{figure*}
\includegraphics[width=8cm]{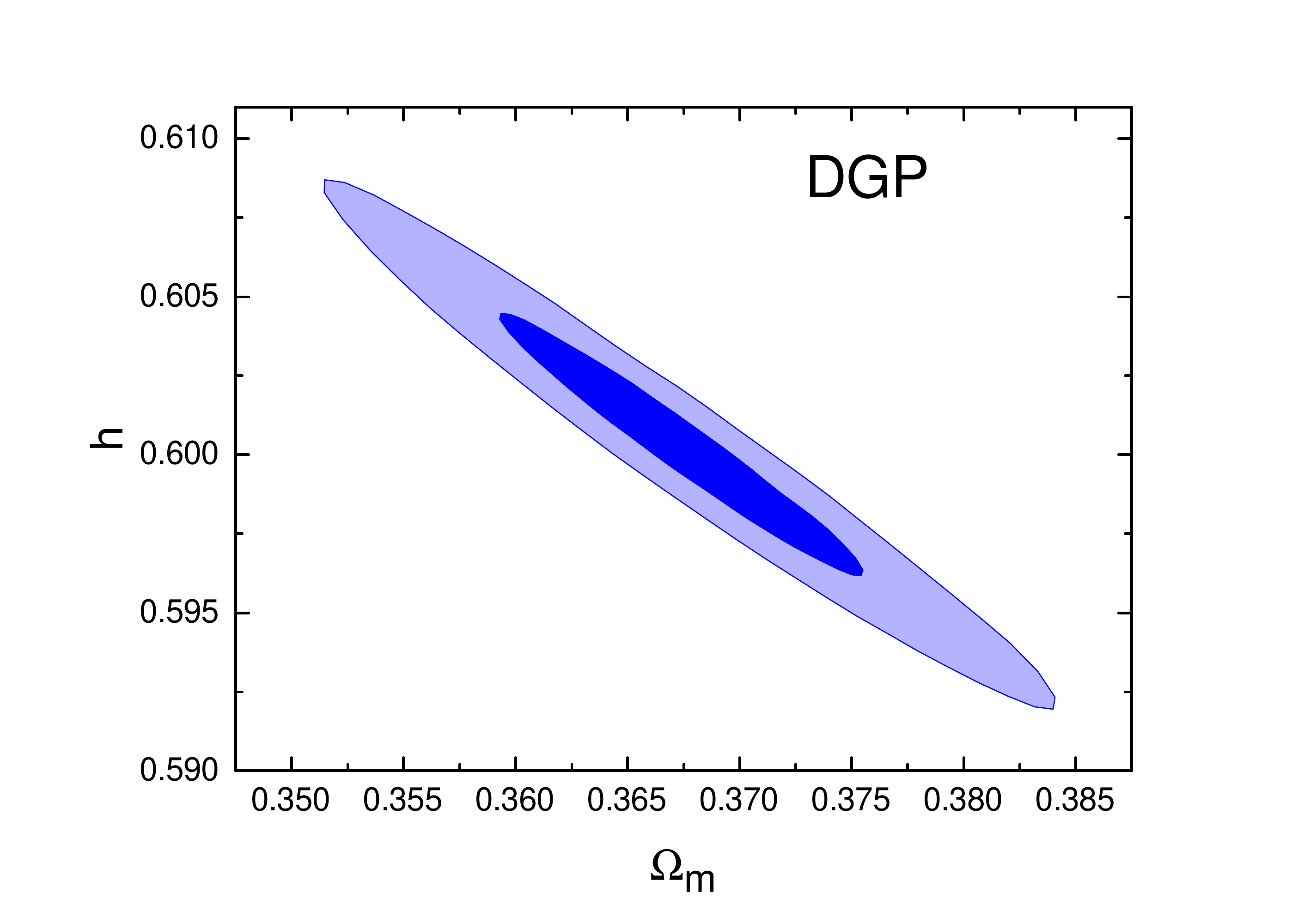}\\
\caption{\label{fig9}The Dvali-Gabadadze-Porrati model: 68.3\% and 95.4\% confidence level contours in the $\Omega_{\rm{m}}$--$h$ plane.}
\end{figure*}

\subsubsection{$\alpha$ dark energy model }

The $\alpha$DE model \cite{c:2003} is a phenomenological extension of the DGP model, in which the Friedmann equation is modified as
\begin{equation}
3M^{2}_{\rm{pl}}\left(H^{2}-\frac{H^{\alpha}}{r^{2-\alpha}_{\rm{c}}}\right)=\rho_{\rm{m}}(1+z)^{3}+\rho_{\rm{r}}(1+z)^{4},
\end{equation}
where $\alpha$ is a phenomenological parameter and $r_{\rm{c}}=(1-\Omega_{\rm{m}}-\Omega_{\rm{r}})^{1/(\alpha-2)}H^{-1}_{0}$. In this model, $E(z)$ is given by the equation
\begin{equation}
E(z)^2=\Omega_{\rm{m}}(1+z)^{3}+\Omega_{\rm{r}}(1+z)^{4}+E(z)^{\alpha}(1-\Omega_{\rm{m}}-\Omega_{\rm{r}}).
\end{equation}
The $\alpha$DE model with $\alpha=1$ reduces to the DGP model and with $\alpha=0$ reduces to the $\Lambda$CDM model. 

From the joint observational constraints, we get the best-fit parameters and the corresponding $\chi^{2}_{\rm{min}}$:\begin{equation}
\Omega_{\rm{m}}=0.326, \;   \alpha=0.106, \;  h=0.663, \;  \chi^{2}_{\rm{min}}=698.574.
\end{equation}
The likelihood contours for the $\alpha$DE model in the $\Omega_{\rm{m}}$--$\alpha$ and $\Omega_{\rm{m}}$--$h$ planes are shown in Fig.~\ref{fig10}.

We find that the $\alpha$DE model performs well in fitting the current observational data. From Fig.~\ref{fig10}, we explicitly see that the DGP limit ($\alpha=1$) is excluded by the current observations at high statistical significance, and the $\Lambda$CDM limit ($\alpha=0$) is well consistent with the current data within the 1$\sigma$ range. For the $\alpha$DE model, we have $\Delta {\rm AIC}=1.199$ and $\Delta {\rm BIC}=5.816$.

\begin{figure*}
\includegraphics[width=8cm]{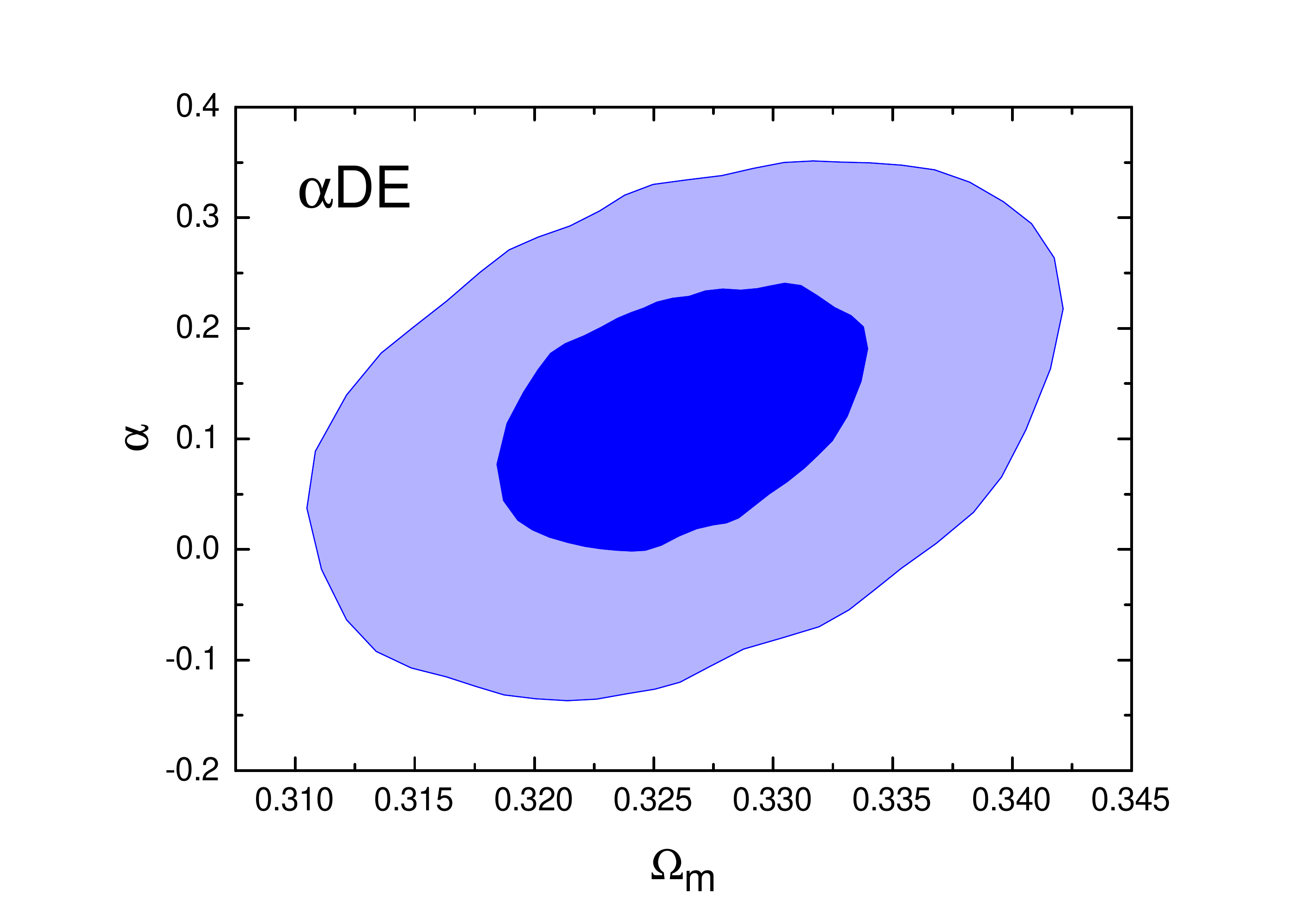}
\includegraphics[width=8cm]{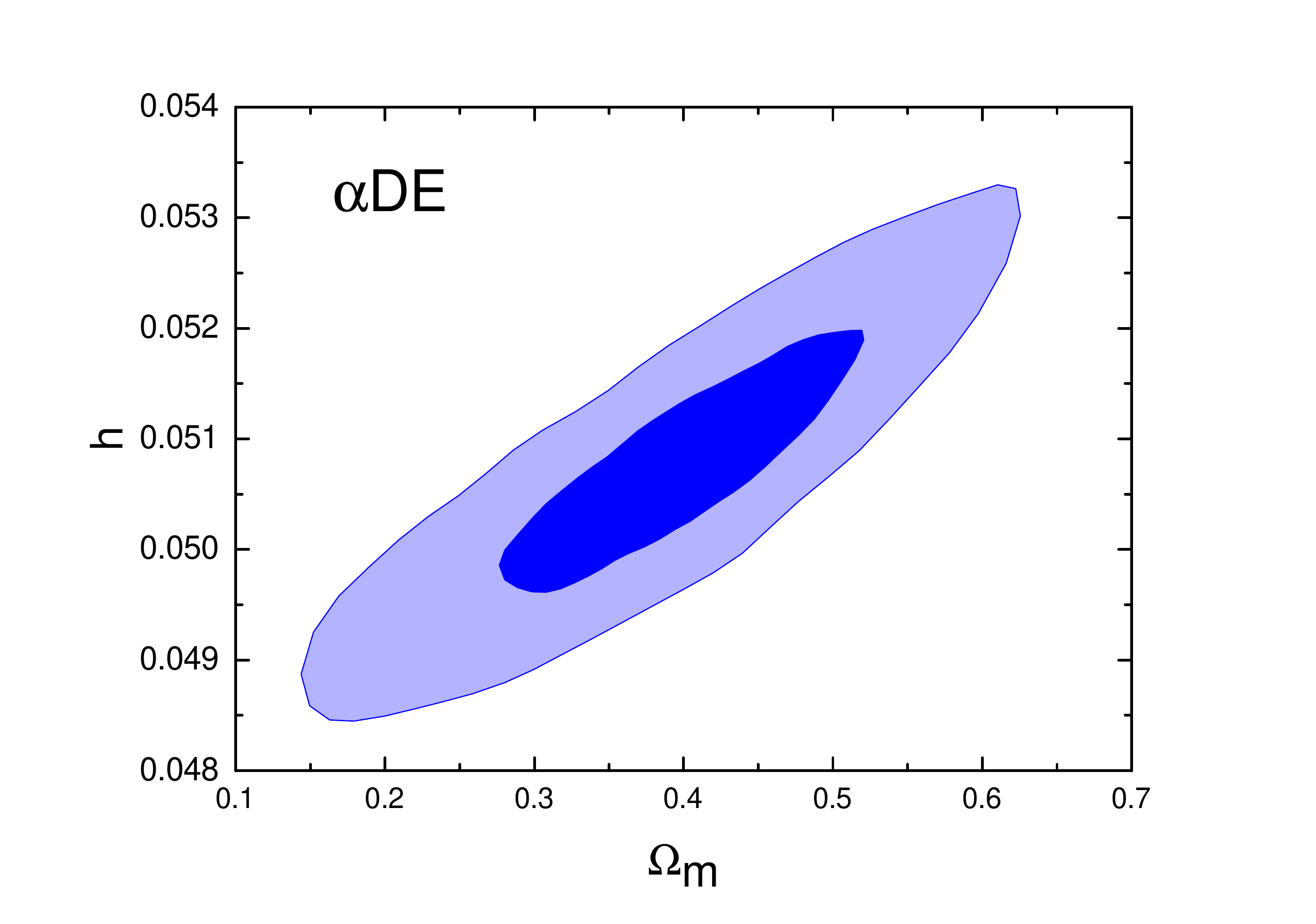}\\
\caption{\label{fig10}The $\alpha$ dark energy model: 68.3\% and 95.4\% confidence level contours in the $\Omega_{\rm{m}}$--$\alpha$ and $\Omega_{\rm{m}}$--$h$ planes.}
\end{figure*}

\section{Discussion and conclusion}\label{sec.5}

\begin{figure*}
\includegraphics[width=10cm]{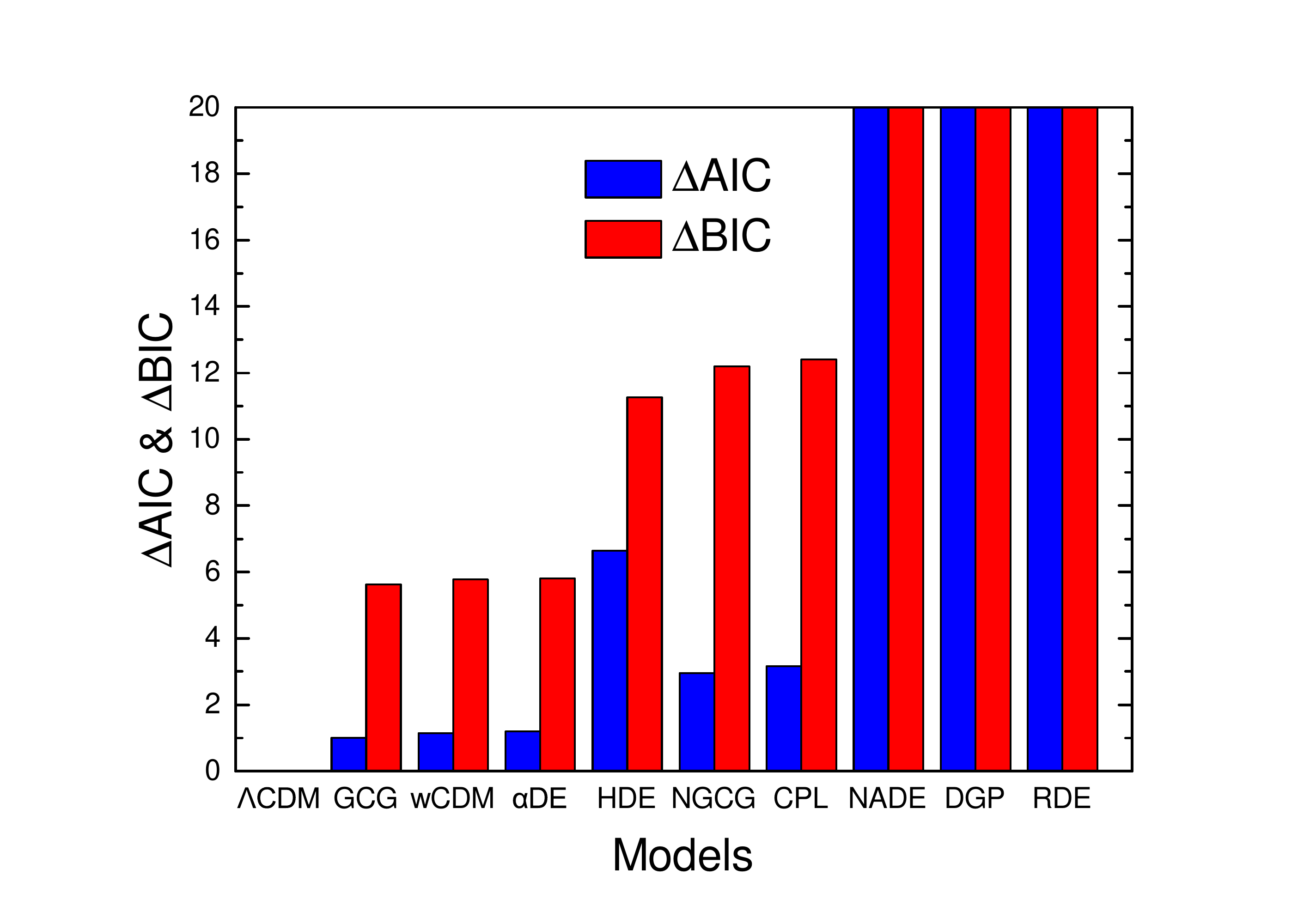}\\
\caption{\label{fig11} Graphical representation of the model comparison result. The order of models from left to right is arranged according to the values of $\Delta{\rm BIC}$, i.e., in order of increasing $\Delta{\rm BIC}$.}
\end{figure*}

We have considered ten typical, popular dark energy models in this paper, which are the $\Lambda$CDM, $w$CDM, CPL, GCG, NGCG, HDE, NADE, RDE, DGP, and $\alpha$DE models. To investigate the capability of fitting observational data of these models, we first constrain these models using the current observations and then make a comparison for them using the information criteria. The current observations used in this paper include the JLA sample of SN Ia observation, the Planck 2015 distance priors of CMB observation, the BAO measurements, and the $H_0$ direct measurement. 

The models have different numbers of parameters. We take the $\Lambda$CDM model as a reference. The NADE and DGP models have the same number of parameters as $\Lambda$CDM. The $w$CDM, GCG, HDE, RDE, and $\alpha$DE models have one more parameter than $\Lambda$CDM. The CPL and NGCG models have two more parameters than $\Lambda$CDM. To make a fair comparison for these models, we employ AIC and BIC as model-comparison tools. 

The results of observational constraints for these models are given in Table \ref{table1} and the results of the model comparison using the information criteria are summarized in Table \ref{table2}. To visually display the model-comparison result, we also show the results of $\Delta {\rm AIC}$ and $\Delta {\rm BIC}$ of these model in Fig. \ref{fig11}. In Table \ref{table2} and Fig. \ref{fig11}, the values of $\Delta {\rm AIC}$ and $\Delta {\rm BIC}$ are given by taking $\Lambda$CDM as a reference. The order of these models in Table \ref{table2} and Fig. \ref{fig11} is arranged according to the values of $\Delta {\rm BIC}$.

These results show that, according to the capability of fitting the current observational data, the $\Lambda$CDM model is still the best one among all the dark energy models. The GCG, $w$CDM, and $\alpha$DE models are still relatively good models in the sense of explaining observations. The HDE, NGCG, and CPL models are relatively not good from the perspective of fitting the current observational data in an economical way. We can confirm that, in the sense of explaining observations, according to our analysis results, the NADE, DGP, and RDE models are excluded by current observations. In the models considered in this paper, only the HDE, NADE, RDE, and DGP models cannot reduce to $\Lambda$CDM, and among these models the HDE model is still the best one. Compared to the previous study \cite{Zhang.xin:2010}, the basic conclusion is not changed; the only subtle difference comes from the concrete orders of models in each group of the above three groups. 

In conclusion, according to the capability of explaining the current observations, the $\Lambda$CDM model is still the best one among all the dark energy models. The GCG, $w$CDM, and $\alpha$DE models are worse than $\Lambda$CDM, but still are good models compared to others. The HDE, NGCG, and CPL models can still fit the current observations well, but from the perspective of providing an economically feasible way, they are not so good. The NADE, DGP, and RDE models are excluded by the current observations.

\acknowledgments
We thank Jing-Lei Cui, Lu Feng, Yun-He Li, and Ming-Ming Zhao for helpful discussions.
This work is supported by the Top-Notch Young Talents Program of China, the National Natural Science Foundation of China (Grant No.~11522540), and the Fundamental Research Funds for the Central Universities (Grant No. N140505002).

\end{document}